\documentclass[letter,11pt]{article}
\usepackage[utf8]{inputenc}

\usepackage{jheppub}
\usepackage{tikz}
\usetikzlibrary{decorations.pathmorphing}
\usetikzlibrary{decorations.markings}
\usepackage{xspace}
\usepackage{booktabs}

\usepackage{amsmath}
\usepackage{amssymb}
\usepackage{mathtools,bm}
\usepackage{empheq}
\usepackage{physics}
\usepackage{slashed}
\usepackage{hyperref}
\usepackage[capitalize]{cleveref}

\DeclareRobustCommand{\parenbar}[1]{%
  \mathchoice%
  {\displaystyle\rlap{$\displaystyle\bar{#1}$}\raisebox{-0.25ex}{$\displaystyle\overset{\displaystyle\kern0.15em\scalebox{.2}{\boldmath$\displaystyle($}\kern0.5em\scalebox{.2}{\boldmath$\displaystyle)$}}{\displaystyle\phantom{#1}}$}}
  {\textstyle\rlap{$\textstyle\bar{#1}$}\raisebox{-0.25ex}{$\textstyle\overset{\textstyle\kern0.15em\scalebox{.2}{\boldmath$\textstyle($}\kern0.5em\scalebox{.2}{\boldmath$\textstyle)$}}{\textstyle\phantom{#1}}$}}
  {\scriptstyle\rlap{$\scriptstyle\bar{#1}$}\raisebox{-0.18ex}{$\scriptstyle\overset{\scriptstyle\kern0.13em\scalebox{.2}{\boldmath$\scriptstyle($}\kern0.35em\scalebox{.2}{\boldmath$\scriptstyle)$}}{\scriptstyle\phantom{#1}}$}}
  {\scriptscriptstyle\rlap{$\scriptscriptstyle\bar{#1}$}\raisebox{-0.14ex}{$\scriptscriptstyle\overset{\scriptscriptstyle\kern0.12em\scalebox{.2}{\boldmath$\scriptscriptstyle($}\kern0.28em\scalebox{.2}{\boldmath$\scriptscriptstyle)$}}{\scriptscriptstyle\phantom{#1}}$}}
\xspace}

\DeclareRobustCommand{\re}{\mathrm{e}} 

\title{Infrared-safe flavoured anti-$k_T$ jets}

\author[a]{Michal Czakon,}
\author[b]{Alexander Mitov,}
\author[b]{Rene Poncelet}

\affiliation[a]{Institut f\"ur Theoretische Teilchenphysik und Kosmologie, RWTH Aachen University, D-52056 Aachen, Germany}
\affiliation[b]{Cavendish Laboratory, University of Cambridge, Cambridge CB3 0HE, United Kingdom}

\emailAdd{mczakon@physik.rwth-aachen.de}
\emailAdd{adm74@cam.ac.uk}
\emailAdd{poncelet@hep.phy.cam.ac.uk}

\note{Preprint: Cavendish-HEP-22/06, P3H-22-056, TTK-22-17}

\abstract{Flavour tagging is technically challenging on the experimental side. However, it suffers from a more fundamental problem from the theoretical point of view, in particular when implemented in fixed-order perturbation theory. It turns out that an infrared-safe definition of a flavoured jet is intricate due to the singularities induced by the emission of flavoured quark-anti-quark pairs of negligible energy. Although this issue has been addressed by a modification of the standard $k_T$ jet algorithm, the situation is not entirely satisfactory as most measurements rather use the anti-$k_T$ jet algorithm. In this work, we propose a flavour-aware infrared-safe modification of the anti-$k_T$ jet algorithm that is easy to implement within perturbative Monte Carlo frameworks and has minor impact on jet phenomenology when flavour tagging is not required. Besides the numerical verification of the infrared safety of the proposed algorithm at next-to-next-to-leading order, we also present results for the hadro-production of a lepton pair in association with a $b$-jet, and of a top-quark pair decaying into $b$-jets and leptons.}

\begin{document} 
\maketitle
\flushbottom

\section{Introduction}\label{sec:intro}

Jets\footnote{For a classic review see Ref.~\cite{Salam:2010nqg}.} are a staple of the research program at high-energy hadron colliders. As suitably defined sets of highly-energetic particles, they constitute a useful tool to establish a link between Quantum Chromodynamics (QCD) of quarks and gluons and the realm of actual strongly-interacting particles, baryons and mesons.

Besides the general importance of jets for collider phenomenology, there is a growing interest in studying jet substructure in order to disentangle various QCD effects governing jet dynamics \cite{Larkoski:2017jix, Marzani:2019hun}. One particular aspect concerns the separation of jets into different categories defined by the underlying partons. For example, the tagging of quark and gluon jets \cite{Gras:2017jty} has received substantial attention. Similarly, jets of definite flavour discussed in the present work, allow to access the partonic structure of the hard scattering event, which in turn can be used for Standard Model measurements, Parton Distribution Function fits, New Physics searches and tuning/improvement of Monte Carlo simulations.

Experimentally, the definition of $b$- or $c$-flavoured jets, i.e.\ the {\it flavour tagging} of jets, is based on the presence, in a given event, of intermediate heavy long-lived $B$- or $D$-hadrons decaying at {\it displaced vertices}, and their association to a jet according to kinematically-defined criteria. A similar procedure can be pursued within Monte Carlo event and/or detector simulations thanks to the resummation of multi-parton radiation effects and hadronisation modelling. On the other hand, obtaining predictions for corresponding observables using partonic computations at fixed-order of perturbation theory in QCD is not as straightforward. Firstly, a prescription that relates partons within a jet to the flavour tag must be introduced. To this end, one assumes that a $b$- or $c$-quark within a jet is related to a $b$- or $c$-jet tag based on $B$- or $D$-hadrons. Secondly, the disparity of energy scales between the collision energy and the mass of flavoured quarks calls for resummation of large logarithms of the ratio of the two scales. This resummation is typically achieved by considering the massless limit and using Renormalisation Group techniques.  Unfortunately, flavoured massless quarks might introduce unphysical singularities if the flavour of hard jets is not correctly determined. The assignment of flavour is non-trivial due to the occurrence of pairs consisting of a quark and anti-quark of the same flavour and vanishingly small energy. In the limit of collinear momenta of the quarks belonging to such a pair, any sensible flavour definition would correctly count the pair as flavourless, since the partons cannot be resolved based on angular separation. On the other hand, if the momenta of the partons have a large angular separation, a careless treatment might lead to hard jets acquiring or losing flavour due to association with only one of the flavoured quarks. This results in an infrared (IR) divergence in the case of flavour-sensitive observables calculated in fixed-order perturbation theory, because the flavour of the pair is resolved while its constituents cannot be observed separately. This problem, originally discussed in Ref.~\cite{Banfi:2006hf}, first occurs at the next-to-next-to leading order (NNLO) of perturbation theory, where two simultaneously-soft partons are allowed. We stress that resummation resolves the issue, since in actuality all soft radiation is exponentially suppressed and states with dangerous incorrect flavour assignment have a negligible contribution to the cross section. This is a prime case where fixed-order perturbation theory is a good model if and only if the observable is IR safe. This IR safety can be practically achieved by using an IR-safe flavour-aware jet definition

The first flavour-aware infrared-safe jet algorithm has been proposed in Ref.~\cite{Banfi:2006hf}. It consists of a modification of the $k_T$ jet algorithm, see Section~\ref{sec:algo}, and has been adopted in a multitude of studies of flavoured-jet production starting from Refs.~\cite{Banfi:2007gu, Banfi:2010xy} (see also \cite{Banfi:2016zlc}) by the authors of the original proposal. Independent applications of the flavoured $k_T$ algorithm at NNLO QCD began with Ref.~\cite{Weinzierl:2006yt} concerned with the forward-backward asymmetry at electron-positron colliders. Another asymmetry defined with the help of the new algorithm was studied in the context of muon decay in Ref.~\cite{Caola:2014daa}. Recently, several pertinent hadron-collider analyses have appeared, in particular for processes with the following final states: $V+H(\to b\bar{b})$ \cite{Caola:2017xuq,Ferrera:2017zex,Gauld:2019yng,Gauld:2021ule} (see also Ref.~\cite{Behring:2020uzq} for comparison with predictions using a finite $b$-quark mass), $Z+b$-jet \cite{Gauld:2020deh}, $W+c$-jet \cite{Czakon:2020coa}, $W+b$ and $\bar{b}$-jets \cite{Hartanto:2022qhh}. The algorithm has also been used in the context of $b$-quark fragmentation into a $B$-hadron to construct a proxy of a fragmentation function in Ref.~\cite{Czakon:2021ohs}.

Although not needed for infrared-finite predictions, the flavoured $k_T$ algorithm has also been used at next-to-leading order (NLO) QCD in Ref.~\cite{Anger:2017glm} concerned with $Wb\bar{b}$ production in association with light (unflavoured) jets. Of course, the lack of singularities at NLO with jets defined with flavour-blind algorithms does not mitigate the fact that singularities would appear at higher orders for the same process. Hence, using a jet algorithm that guarantees sensible results to all orders is a well-motivated strategy. Similarly, although cross section predictions including resummation of multi-parton emissions do not require the use of a flavour-aware jet algorithm, several studies made use of the flavoured $k_T$ algorithm in the context of parton showers \cite{Astill:2018ivh, Bizon:2019tfo, Zanoli:2021iyp, Bevilacqua:2021tzp} and analytic resummation \cite{Larkoski:2013eya, Baberuxki:2019ifp}.

Regardless of these successes, Ref.~\cite{Banfi:2007gu} points out a major problem in using the flavoured $k_T$ algorithm when performing comparisons with data: the algorithm is defined in terms of partons and cannot be trivially translated to hadron-based events. The main issue is the limited efficiency of flavour tagging (correct identification of a jet as originating from a flavoured quark) and non-vanishing probability of mis-tagging (incorrect identification of a jet originating from non-flavoured partons as flavoured). The problem is exacerbated by the necessity to determine the charge of the original flavoured quarks within the hadrons, which introduces its own uncertainties. Ref.~\cite{Banfi:2007gu} discusses the consequences of these factors within a simplified model, and concludes with the statement that the algorithm could, in principle, be applied in experimental analyses at the cost of additional systematic uncertainties. Despite the constant progress in flavour tagging since the publication of Ref.~\cite{Banfi:2007gu}, imperfect tagging efficiency, mis-tagging and charge identification inefficiency remain. Hence, even if the algorithm were applied at the hadron level in the experimental setting, it would not match one-to-one onto the same algorithm applied at the parton level in the theoretical calculation. The fact that the experimental collaborations did not adapt the flavoured $k_T$ algorithm is, however, a consequence of lack of motivation. Indeed, the usage of a flavour-blind jet algorithm with subsequent flavour tagging would still be necessary in practice, see details in Ref.~\cite{Banfi:2007gu}. The flavour-aware jet algorithm would have to be applied in a second step. This procedure is very similar to {\it unfolding}.

In order to compensate for the difference between the flavoured $k_T$ algorithm and the flavour tagging in experiment, it is necessary to {\it unfold} the measurement data, i.e.\ translate from experimental to theory definitions. This is achieved by comparing the results of parton shower simulations employing two different jet definitions: 1) flavoured-$k_T$ to match the theoretical predictions; 2) standard anti-$k_T$ with usual flavour tags to match experimental measurements. This procedure has been studied carefully in Ref.~\cite{Gauld:2020deh}, and yields in that particular case a $\sim 10\%$ change of $b$-jet distributions. Due to the inherent differences between the $k_T$ and anti-$k_T$ algorithms the effect in other processes might be further enhanced.

Alternatives to the flavoured $k_T$ algorithm are rare. One approach consists of using the flavoured $k_T$ algorithm to determine the flavour of otherwise anti-$k_T$-clustered \cite{Cacciari:2008gp} jets. This has been tried at NLO in Ref.~\cite{Heinrich:2013qaa} and in the context of resummation in Ref.~\cite{Caletti:2021oor}. A more sophisticated flavour assignment along similar lines has been proposed in Ref.~\cite{Buckley:2015gua}. It is also worth mentioning that machine learning techniques have also been explored as an alternative, see Ref.~\cite{Fedkevych:2022mid}. The latest proposal of an alternative for flavour identification has appeared while the present publication was being prepared \cite{Caletti:2022glq,Caletti:2022hnc}. Here, in an extension of Ref.~\cite{Caletti:2021oor}, it is proposed to use soft-jet grooming for flavour tagging.

In the present publication, we propose a modification of the anti-$k_T$ algorithm with the following properties: 1) it renders jet-flavour infrared safe; 2) it is easy to implement in a fixed-order Monte Carlo calculation; 3) it only leads to slight differences with respect to predictions based on the standard anti-$k_T$ algorithm for non-flavour-sensitive observables.

This work is structured as follows. In Section \ref{sec:algo}, we describe the proposed flavour-aware infrared-safe modification of the anti-$k_T$ algorithm. In Section \ref{sec:irsafety}, we illustrate the infrared safety of the algorithm with two different numerical methods. In the following Section \ref{sec:pheno}, we present phenomenological results for the processes $pp \to Z/\gamma^*(\to \ell\bar{\ell}) + b$-jet and $pp \to t\bar{t} \to \ell\bar{\ell}\nu\bar{\nu} + 2 b$-jets. We close in Section \ref{sec:conclusion} with a discussion and outlook. Unfolding corrections in the case of $pp \to Z/\gamma^*(\to \ell\bar{\ell}) + b$-jet are provided in an Appendix.

\section{The jet algorithm}
\label{sec:algo}

\subsection{Infrared safety and flavour}

Before discussing jet flavour, let us first recapitulate the fundamental properties of infrared-safe jet algorithms.

A jet algorithm clusters partons, if applied within QCD, or particles, if applied at the level of physical final states. Infrared safety is a \emph{sine qua non} requirement for the former case, and can be stated as follows: the quantitative properties of a jet, in particular its four-momentum, must not change in case its members emit additional particles of negligible energy (\emph{soft emissions}) or split into several particles travelling in nearly the same direction (\emph{collinear emissions}). If this condition is fulfilled then soft and/or collinear singularities of virtual and real corrections in a fixed-order perturbative cross section calculation cancel each other at the level of jets. 

All algorithms used in practice at hadron colliders\footnote{Prior to the operation of the LHC, cone-based algorithms have also been in use, for a review see Ref.~\cite{Salam:2010nqg}.} consist of iterative two-to-one recombination steps based on a distance measure. Each iteration merges two elements with the smallest distance out of a list of \emph{pseudo-jets} initialised with the final state particles. The algorithm proceeds until a stopping condition is fulfilled. For \emph{exclusive} algorithms this condition amounts to requiring a given number of jets. \emph{Inclusive} algorithms, on the other hand, remove pseudo-jets that fulfill a certain property until the list is empty. Jet algorithms rely on an angle-dependent distance measure that vanishes for collinear pseudo-jet momenta in order to capture the singular scaling of QCD amplitudes in the collinear limit. The handling of soft emissions varies substantially between algorithms, because there is no need to cluster soft emissions in order to ensure infrared safety, as long as the only relevant jet property is its four-momentum. Indeed, an emission at vanishing energy, hence also four-momentum, would not change a jet's four-momentum anyway. Nevertheless, the dependence of the distance measure on pseudo-jet's energy leads to different recombination patterns. That, in turn, may yield observables that have attractive properties as far as phenomenological applications are concerned. This is, for instance, the reason for the prevalent use of the anti-$k_T$ jet algorithm in comparison to the older $k_T$ jet algorithm.

Let us now turn to jet flavour. Infrared safety demands that this quantum number, just as jet momentum, be constant in all infrared limits. In a fixed-order cross section calculation, the concept of jet flavour relies on the flavour of the clustered partons. As usual, we define it to be the difference of the number of flavoured quarks and that of flavoured anti-quarks, possibly modulo two. It is understood that charm (anti-)quarks and bottom (anti-)quarks are counted separately, which yields values for $c$- and $b$-flavour respectively. Clearly, there is a crucial difference between momentum and flavour: flavour is a discrete quantity. In consequence, while arbitrary emissions of soft pairs of flavoured quarks and anti-quarks do not change the momentum of a jet in the vanishing-energy limit, they might incorrectly affect its flavour. These emissions are predominantly generated by the splitting of virtual gluons which results in a singularity of the cross section at fixed order of perturbative QCD starting from NNLO. The singularity is cancelled by virtual corrections assuming that both real and virtual contributions yield the same value of the studied observable in the strict soft limit. This requirement must be taken into account by the jet algorithm. Jet flavour is clearly only then infrared safe if soft flavoured quark-anti-quark pairs of zero net flavour are clustered together. This condition establishes a link between the so-far unrelated concepts of kinematics and flavour. For non-flavour-aware standard jet algorithms, infrared safety may only be violated by soft flavoured quarks and anti-quarks that are far from collinear to each other. In the collinear case, they would be clustered together irrespective of the energy. For wide-angle emissions, it may happen that the two partons are clustered within two different jets altering their flavour in the process. These real-radiation contributions will then potentially yield a different value of a given flavour-sensitive observable compared to the value for the contribution with the pair being virtual, and there would remain an uncancelled divergence.

We now formulate criteria for infrared safety in the presence of flavour tagging. Let $d_{ij}$ be a distance measure between the pseudo-jets $i$ and $j$. Furthermore, let the pseudo-jets have energies $E_i,E_j$, transverse momenta $k_{T,i},k_{T,j}$, rapidities $y_i,y_j$ and azimuths $\phi_i,\phi_j$. The angular separation is defined as:
\begin{equation}
    R^2_{ij} \equiv \Delta R^2_{ij}/R^2 \; , \qquad \Delta R^2_{ij} \equiv (y_i-y_j)^2+(\phi_i-\phi_j)^2 \; ,
\end{equation}
with $R$ an arbitrary parameter called the \emph{jet radius}. Infrared safety is ensured if the following two conditions are met in the wide-angle double-soft limit, $E_i,E_j \to 0$ with $i$ and $j$ of opposite-sign flavour:
\begin{enumerate}
  \item[1)] $d_{ij}$ vanishes for arbitrary $R_{ij}$;
  \item[2)] $d_{ij}$ vanishes faster than the distance of either $i$ or $j$ to the remaining pseudo-jets.
\end{enumerate}
Both conditions are necessary, since otherwise a flavoured parton that actually belongs to a soft flavourless pair can be recombined with a hard pseudo-jet, resulting in a hard pseudo-jet of altered flavour. The incorporation of these conditions in existing jet algorithms leads to non-trivial modifications. Below we first recall the flavoured $k_T$ jet algorithm introduced in Ref.~\cite{Banfi:2006hf}, and then present our proposal for a flavoured anti-$k_T$ jet algorithm.

\subsection{The flavoured $k_T$ algorithm}

The distance measure of the standard $k_T$ algorithm is:
\begin{equation}
  d_{ij} =  R^2_{ij} \, \min (k_{T,i}^2,k_{T,j}^2) \,.
\end{equation}
It already fulfills condition 1), since it vanishes in the double-soft limit independently of the distance $R_{ij}$. Condition 2), however, is not fulfilled since the minimum of the two transverse momenta is used, which yields the same numerical value for soft-soft and soft-hard pairings, i.e.\ the transverse momentum of the softer pseudo-jet. The algorithm is made infrared safe by the following modified distance measure \cite{Banfi:2006hf}:
\begin{equation}
	\label{eq:flavourkt_meas}
	d^{(F)}_{ij} = R_{ij}^2 \times
	\begin{cases}
		\big[ \max(k_{T,i}, k_{T,j}) \big]^{\alpha}\,\big[\min(k_{T,i}, k_{T,j})\big]^{2-\alpha} \, , & \text{if softer of } i, j \text{ is flavoured,} \\
		\min(k^2_{T,i}, k^2_{T,j}) \, , & \text{if softer of } i, j \text{ is unflavoured,} \\
	\end{cases}
\end{equation}
where $0 < \alpha \leq 2$ and most analyses are performed with $\alpha = 2$.
This jet algorithm modification prevents the unwanted soft-hard recombination if the softer pseudo-jet is flavoured, while it still leads to soft-soft recombination. One can additionally require recombination into pseudo-jets of well-defined flavour only, by forbidding, for example, charm and beauty to be recombined. This is the ``bland'' version of the algorithm in Ref.~\cite{Banfi:2006hf}.

Until now, we have ignored initial state radiation and the related singularities. In the standard $k_T$ algorithm, one defines a distance to the beam, $d_{i\parenbar{B}} = k_{T,i}^2$. If it is minimal, then the pseudo-jet $i$ is removed from the list of pseudo-jets in the inclusive formulation. In the flavoured $k_T$ algorithm, the distance to the beam is modified as well. Indeed this is necessary, since if $i$ contains a soft flavoured quark while there is another soft anti-quark of the same flavour that would not be removed from the list, but rather clustered with a hard jet, then infrared safety would be spoiled. The beam distance is thus defined in analogy to the case of final-state pseudo-jets as follows:
\begin{equation}
	\label{eq:flavourkt_beam}
	d^{(F)}_{i\parenbar{B}} =
	\begin{cases}
		\big[\max(k_{T,i}, k_{T,\parenbar{B}}(y_i))\big]^{\alpha}\,\big[\min(k_{T,i}, k_{T,\parenbar{B}}(y_i))\big]^{2-\alpha} \, , & \text{if } i \text{ is flavoured,}   \\
		\min(k^2_{T,i}, k^2_{T,\parenbar{B}}(y_i)) \, , & \text{if } i \text{ is unflavoured.} \\
	\end{cases}
\end{equation}
The now required transverse momentum of the beam, $B$, and ``anti-''beam, $\bar{B}$, is taken to be \cite{Banfi:2006hf}:
\begin{align}
	k_{T,B}(y)       & = \sum_i k_{T,i} \left( \Theta(y_i - y) +  \Theta(y - y_i) \; \re^{y_i-y} \right), \label{eq:ktb}    \\
	k_{T,\bar{B}}(y) & = \sum_i k_{T,i} \left( \Theta(y - y_i) +  \Theta(y_i-y)   \; \re^{y-y_i} \right), \label{eq:ktbbar}
\end{align}
with $ \Theta(0) = 1/2$.

\subsection{The flavoured anti-$k_T$ algorithm}

The distance measure of the standard anti-$k_T$ algorithm \cite{Cacciari:2008gp} is:
\begin{equation}
  d_{ij} =  R^2_{ij} \, \min (k_{T,i}^{-2},k_{T,j}^{-2}) \,.
  \label{eq:anti-kt}
\end{equation}
In this case, condition 1) is not fulfilled, since the double-soft limit, $E_i,E_j \to 0$, does not lead to a vanishing $d_{ij}$. We propose the following modification:
\begin{empheq}[box=\fbox]{equation} \label{eq:flavour-anti-kT}
  d_{ij}^{(F)} \equiv d_{ij} \times
  \begin{cases}
   \mathcal{S}_{ij} \, , & \text{if both $i$ and $j$ have non-zero flavour of opposite sign,} \\[.2cm] 
   1 \, , & \text{otherwise.}
  \end{cases}
\end{empheq}
In the double-soft limit, the \emph{damping function} $\mathcal{S}_{ij}$ is defined to vanish faster than $E^2$ in order to overcome the scaling $d_{ij} \sim 1/E^2$ $(E \to 0)$, where $E$ is the energy of the harder of the two soft flavoured (anti-)quarks. With this assumption, soft flavoured quark-anti-quark pairs will be clustered before anything else in the double-soft limit with otherwise fixed angles, as only $d_{ij}$ of such pairs vanishes in that case. In other words, both conditions 1) and 2) are fulfilled. One may still wonder, what happens if the angular separation between a flavoured (anti-)quark and another parton vanishes as well. Assume that $\mathcal{S}_{ij} \sim E^n$ $(E \to 0)$, $n > 2$. The unwanted clustering of flavoured (anti-)quark $i$ with a pseudo-jet $k$ will take precedence if $R_{ik}^2 \; \propto \; \theta_{ik}^2 < \mathcal{C} E_i^{n-2}$ with $\theta_{ik}$ the angle between the momenta of $i$ and $k$, and for some kinematics-dependent $\mathcal{C}$. The probability of a single-collinear emission corresponding to the unwanted clustering is proportional to $1/\theta_{ik}$. It is, nevertheless, not singular in $E_i$. Hence, the resulting logarithmic collinear enhancement is linearly suppressed by $E_i$ and therefore not singular in the double-soft limit where $E_i \to 0$. In consequence, we can neglect this kind of configurations in the discussion of infrared safety.

In order to complete the definition of the flavoured anti-$k_T$ algorithm, we propose the following damping function that yields $d^{(F)}_{ij} \sim E^2$ $(E \to 0)$ as in the flavoured $k_T$ algorithm:
\begin{empheq}[box=\fbox]{equation} \label{eq:Sij}
 \mathcal{S}_{ij} \equiv 1-\theta\left(1-\kappa_{ij}\right)\cos\left(\frac{\pi}{2}\kappa_{ij}\right)
 \quad \text{with} \quad \kappa_{ij} \equiv \frac{1}{a} \, \frac{k_{T,i}^2+k_{T,j}^2}{2 k_{T,\text{max}}^2}\; .
\end{empheq}
$k_{T,\text{max}}$ is set to the  $k_T$ of the hardest pseudo-jet in a clustering step, but other choices are possible as well. For example, it could be taken equal to the renormalisation scale as long as the latter is not jet based. The parameter $a$, on the other hand, can be used to steer the turn-on of the damping function. We discuss how to choose its value in the next section.

According to the previous discussion, $S_{ij}$ guarantees IR safety to all orders if it is defined through the energies of the partons and has appropriate scaling in the double-soft limit. However, Eq.~\eqref{eq:Sij} defines $S_{ij}$ through transverse momenta. Since the latter are bounded by energies from above, they must also vanish in the double-soft limit. Unfortunately, a small transverse momentum does not imply small energy. It is therefore necessary to prove that IR safety is not spoiled due to clustering of soft and hard flavoured-quarks in case the latter are collinear to an initial state. To this order, let us consider a final state with three flavoured quarks, $b(k_i)$, $\bar{b}(k_j)$ and $b(k_k)$, with momenta indicated in the parentheses. We assume that $\bar{b}(k_j)$ and $b(k_k)$ are a soft pair, while $b(k_i)$ is hard-collinear to an initial state. This is the configuration that could endanger IR safety of our jet algorithm, if in the double-soft limit, $E_{k,j} \to 0$, at fixed rapidities $y_{k,j} =$ const, with $E_k \sim E_j \ll E_i$, the quarks $b(k_i)$ and $\bar{b}(k_j)$ are clustered instead of $\bar{b}(k_j)$ and $b(k_k)$. Indeed, in this case, $b(k_k)$ will not be clustered with the flavourless $\big( b(k_i),\bar{b}(k_j) \big)$ system anymore if there is nearby hard jet. Rather, the flavour of this hard jet will be modified by clustering with $b(k_k)$.

We now demonstrate that, with the specified assumptions, $\bar{b}(k_j)$ and $b(k_k)$ will be clustered first. Since $E = k_T \cosh y$, there is $k_{T,k} \sim k_{T_,j}$, where $\sim$ denotes same scaling in the limit. In consequence, by Eqs.~\eqref{eq:anti-kt}, \eqref{eq:flavour-anti-kT} and \eqref{eq:Sij}:
\begin{equation}
    d^{(F)}_{jk} \sim \Delta R^2_{jk} k_{T,j}^2 \; .
\end{equation}
On the other hand:
\begin{equation}
d^{(F)}_{ij} \sim \Delta R^2_{ij} \max(k_{T,i}^2,k_{T,j}^2) \; .
\end{equation}
We assume $k_{T,i} \to 0$, since otherwise $d^{(F)}_{ij}$ cannot be smaller than $d^{(F)}_{kj}$ in the limit. Taking the energy of $b(k_i)$ as reference energy scale to define the double-soft limit, $E_i$ can be considered fixed since $E_k\sim E_j \ll E_i$. Hence $k_{T,i} \sim 1/\cosh y_i$ and $y_i \sim \ln k_{T,i} \to \infty$. This implies in turn an increasing rapidity gap between $b(k_i)$ and $\bar{b}(k_j)$, and thus $\Delta R_{ij}^2 \sim \ln^2 k_{T,i}$. On the other hand since $k_{T,j}$ and $k_{T,k}$ vanish together with their respective energies $E_j$ and $E_k$ we have $\Delta R^2_{jk} \sim 1$. We thus obtain:
\begin{equation}
d^{(F)}_{ij} \sim \ln^2 k_{T,i} \max(k_{T,i}^2,k_{T,j}^2) \; .
\end{equation}
Finally, therefore, independently of the scaling of $k_{T,i}$ with respect to $k_{T,j}$:
\begin{equation}
    d^{(F)}_{jk} / d^{(F)}_{ij} \quad \longrightarrow 0 \; .
\end{equation}
This proves that $\bar{b}(k_j)$ and $b(k_k)$ will always be clustered first in the double-soft limit, and IR safety is not spoiled.

As a final comment, we note that there is no need to modify the distance to the beam, which in the standard anti-$k_T$ algorithm is:
\begin{empheq}[box=\fbox]{equation}
    d_{i\parenbar{B}} = k_{T,i}^{-2} \equiv d^{(F)}_{i\parenbar{B}}\; .
\end{empheq}
With this definition, soft radiation is never clustered to the beam. Hence, there is no danger of incorrect flavour assignment, where both the beam and a hard jet would acquire flavour from soft quarks.

\section{Tests of infrared safety}\label{sec:irsafety}

In this section, we describe two different classes of tests of IR safety of the proposed flavoured anti-$k_T$ algorithm. The first class exploits fixed order computations at NNLO QCD, while the second relies on a parton shower to simulate multiple emissions of flavoured pairs. Both require an IR-sensitive observable to investigate the behaviour of the algorithm.

\subsection{Tests based on cutoff dependence}\label{sec:pheno_cutoff}

We first consider flavour-sensitive jet cross sections, and study different partonic processes contributing to double-real NNLO QCD corrections. We are, in particular, interested in the dependence of the latter on the numerical cutoff parameter\footnote{The exact definition of this parameter is not of importance to the argument and is implementation specific in any case.} $x_\text{cut}$. Such a parameter is present in every numerical code to prevent the evaluation of numerically-unstable cross section contributions at phase space points very close to IR limits. An IR-safe cross section is regular in the limit of vanishing $x_\text{cut}$. In practice, the residual dependence for small but non-zero values of $x_\text{cut}$ is hidden within the Monte Carlo integration-error estimate. A dependence on $x_\text{cut}$ that can be convincingly observed outside of the Monte Carlo uncertainty indicates an IR-unsafe definition of the observable\footnote{This holds assuming that the subtraction scheme used for the calculation has been correctly implemented.}. If the jet algorithm is not IR safe, then there exist jet-based observables that are IR unsafe.

For our investigation, we take the following processes:
\begin{description}
\item[I.] Contributions to the $pp \to Z/\gamma^*(\to \ell \bar{\ell})+b$-jet process:
\begin{enumerate}
    \item $gb \to \ell \bar{\ell} b g g$ (IR-safe flavour structure);
    \item $gb \to \ell \bar{\ell} b b \bar{b}$ (IR-sensitive flavour structure).
\end{enumerate}
\item[II.] Contributions to the $pp \to t\bar{t}$ process including Narrow Width Approximation decays of the top quarks in the di-lepton channel:
\begin{enumerate}
    \item $gg \to t(\to b \bar{\ell} \nu) \bar{t}(\to \bar{b} \ell \bar{\nu}) g g$ (IR-safe flavour structure\footnote{The $b\bar{b}$-pair from top-quark decays does not imply a singular double-soft limit since it cannot result from a soft-gluon splitting.});
    \item $gg \to t(\to b \bar{\ell} \nu) \bar{t}(\to \bar{b} \ell \bar{\nu}) b \bar{b}$ (IR-sensitive flavour structure).
\end{enumerate}
\end{description}
As indicated, we also consider partonic processes for which no jet-based observable is IR sensitive even if the jet algorithm is not IR safe in general. The precise fiducial phase space definitions used in the calculations are stated in Sections \ref{sec:pheno_ppzb} and \ref{sec:pheno_pptt}.

\begin{figure}[t]
 \centering
 \includegraphics[page=1,width=0.45\textwidth]{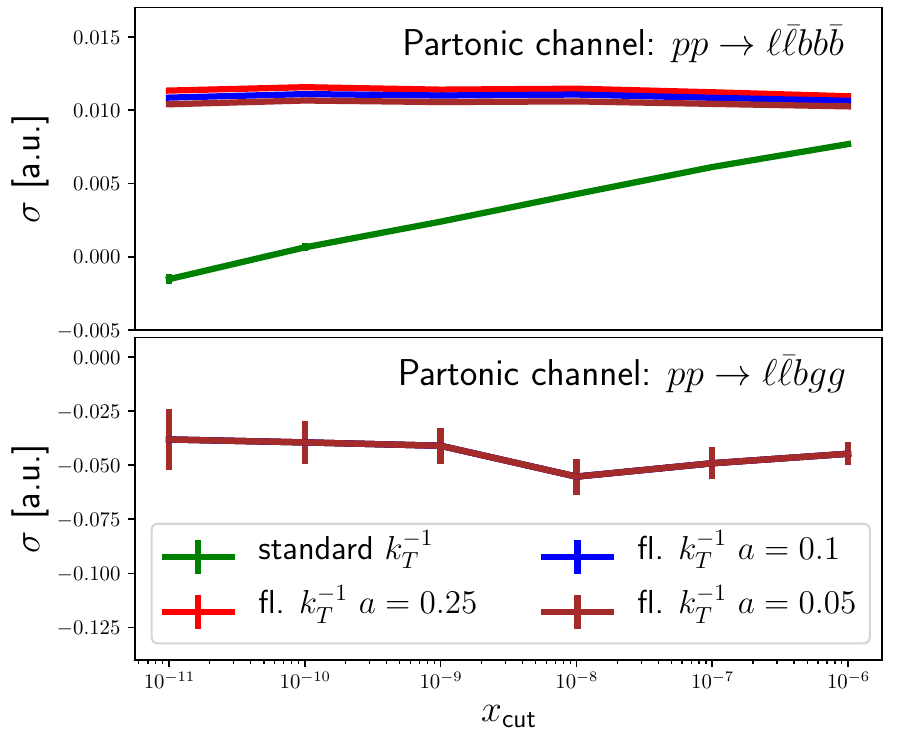}
 \includegraphics[page=2,width=0.45\textwidth]{./figs/irsafety-cutoff.pdf}
 \caption{Cutoff dependence of IR-sensitive (top row) and IR-safe (bottom row) cross sections for different partonic channels contributing to double-real radiation at NNLO QCD for selected processes discussed in the text.}
 \label{fig:cutoff}
\end{figure}

The dependence of the fiducial cross sections on the $x_\text{cut}$ parameter is shown in Fig.~\ref{fig:cutoff}. The left panels correspond to $Z/\gamma^*(\to \ell \bar{\ell})+b$-jet production, while the right panels correspond to top-quark pair production and decay in the di-lepton channel. The plots in the top row depict the contributions that are IR-sensitive in principle, while those of the bottom row should show a stabilisation of the dependence on the cutoff. In each case, we compare the standard anti-$k_T$ algorithm to the proposed flavoured version with different $a$-parameter choices.

Let us discuss contributions with IR-sensitive flavour structure first.
From the top row of Fig.~\ref{fig:cutoff}, it can clearly be seen that the dependence on $x_\text{cut}$ for the IR-safe variant of the anti $k_T$ algorithm is flat in the limit $x_\text{cut} \to 0$, while the standard anti-$k_T$ algorithm demonstrates a logarithmic IR-divergent behaviour. The dependence on the cutoff is so strong in the latter case that the Monte Carlo integration errors are not visible on the scale of the plots. As expected, there is noticeable dependence on the jet algorithm parameter $a$. We analyse it further in the next Section.

In the case of IR-safe flavour structure, there is no divergence at $x_\text{cut} \to 0$ independently of the jet algorithm used as shown in the bottom row of Fig.~\ref{fig:cutoff}. In this case, the size of the Monte Carlo integration error matters. Interestingly,
there is no dependence on the jet algorithm parameter $a$ in the case of $Z/\gamma^*(\to \ell \bar{\ell})+b$-jet production for the chosen partonic channel. This is simply due to the fact that there are no (potentially soft) flavoured pairs in the final state. At variance, a slight dependence is observable for top-quark pair production. Indeed, in this case the modification of the jet algorithm already changes the Born cross section, since the recombination of the $b$ and $\bar{b}$ quarks from the top-quark decays is affected.

\subsection{Tests with showered events}
\label{sec:ir-tests}

Tests along the lines of the previous section, but beyond the case of a single soft flavoured-pair emission would require access to higher orders of perturbation theory. This is, unfortunately, out of reach with the available computational techniques. The emission of multiple soft flavoured pairs can, nevertheless, be simulated using a parton shower. By construction, a leading-log-accurate parton shower yields IR-finite results even for flavour-sensitive jet observables based on IR-unsafe jet algorithms. However, a judiciously constructed observable may at least show enhanced IR sensitivity in the latter case. Observables that are dominated by events close to the Born configuration for a given final state are particularly suited for this purpose. An example of such an observable is the partonic N-jettiness, $\tau_N$, in the process $pp \to N j+X$. For a Born configuration, the value of N-jettiness is $\tau_N = 0$, and there may be $M\leq N$ flavoured jets. Any additional emission yields a non-zero value, $\tau_N > 0$. Close to the endpoint, $\tau_N \approx 0$, IR limits for the additionally emitted partons are enforced. An IR-safe jet algorithm should therefore yield the same number of flavoured jets at $\tau_N = 0$ and for $0 \neq \tau_N \simeq 0$.

\begin{figure}[t]
 \centering
  \includegraphics[page=1,width=0.45\textwidth]{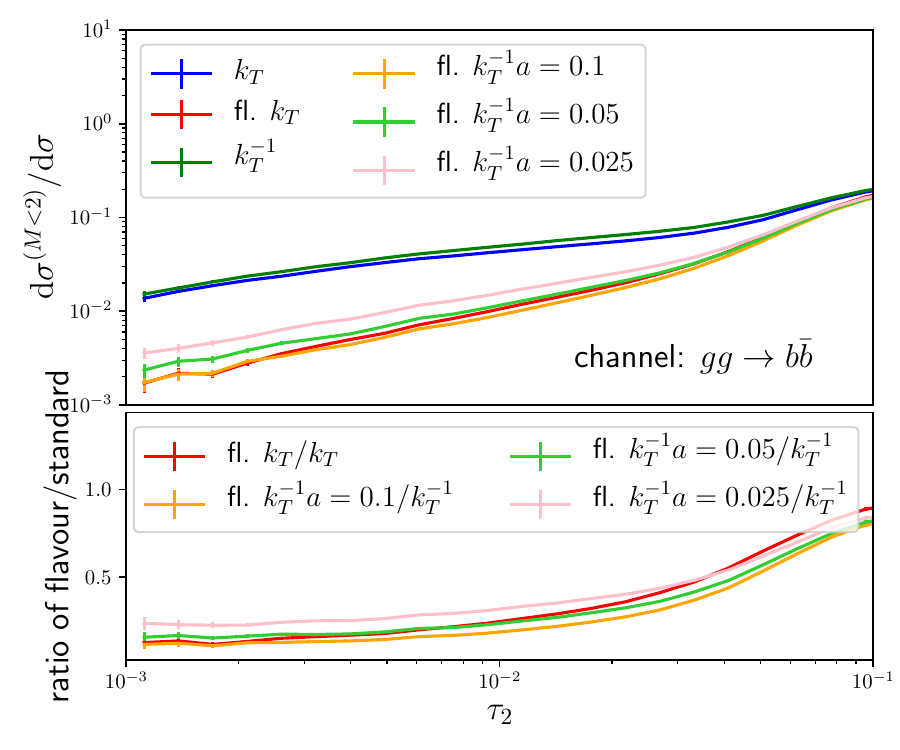}
 \includegraphics[page=2,width=0.45\textwidth]{./figs/irsafety-flakt-v1.pdf}
 \caption{Upper panel: fraction of events with incorrect flavour assignment, $\dd{\sigma^{(M<2)}}/\dd{\sigma}$ and $\dd{\sigma^{(M\neq1)}}/\dd{\sigma}$, for the two leading jets after showering of $gg \to b\bar{b}$ (left column) for $bg \to bg$ (right column) events followed by clustering with various jet algorithms. Lower panel: ratio of the fractions for IR-safe and IR-unsafe jet algorithms.}
 \label{fig:irsafety_ps}
\end{figure}

We demonstrate the above behaviour by considering tree-level events for $gg \to b\bar{b}$ in 13 TeV proton-proton collisions showered using the \textsc{SHERPA} \cite{Gleisberg:2008ta} Monte Carlo framework. We remain at the parton level and do not run the hadronisation stage. We work with massless $u,d,s,c$ and $b$ quarks in the underlying Born events as well as in the shower. We choose a sample of events which have $p_T(j_1) > 1$ TeV, $|\eta(j_1)| < 1$ and $|\eta(j_2)| < 1$ after clustering with the standard $k_T$ algorithm with a radius parameter of $R = 1$. To be explicit there is no flavour assignment required for this sample. For each selected event we compute the partonic quantity $\tau_2$ with respect to the direction of the two leading $k_T$ jets. Finally, we cluster the selected events with different jet algorithms: the standard anti-$k_T$, the flavour $k_T$ as defined in Ref.~\cite{Banfi:2006hf}, and the proposed flavoured anti-$k_T$ jet algorithm. For each set of clustered events we identify the flavour of the two leading jets. If the number of flavoured jets after clustering is not two, as in the underlying Born process, the event is considered to have incorrect flavour assignment. The fraction of such events as a function of $\tau_2$ is expected to decrease for any jet algorithm. However, the rate of decrease must be substantially faster for an IR-safe than for an IR-unsafe algorithm. This behaviour is shown in Fig.~\ref{fig:irsafety_ps} for the algorithms at hand. Clearly, the flavour-aware algorithms perform as expected. We observe a stronger suppression of incorrect flavour assignment with the flavoured anti-$k_T$ algorithm than with the flavoured $k_T$ algorithm in the case of the $gb \to gb$ channel, while the suppression is very similar for both algorithms in the $gg \to b\bar{b}$ channel. These observations further confirm the correctness of our algorithm.

\section{Impact on phenomenology}\label{sec:pheno}

In this section, we present flavour-sensitive differential distributions for two relevant LHC processes with the purpose of demonstrating the properties and features of the proposed flavoured anti-$k_T$ jet algorithm. As a first example, we study the production of a charged lepton pair created by an intermediate $Z/\gamma^*$ gauge boson in association with a $b$-flavoured jet. As a second example, we consider the production of a top-quark pair with subsequent decay into $b$-flavoured jets and leptons. While this process is well-defined without cuts on the final states due to the employed Narrow Width Approximation, the definition of $b$-jets is of relevance when fiducial phase space cuts are in place, or in the case of $b$-jet observables, such as the transverse momentum of the leading $b$-jet. Both analyses are performed in the five-flavour scheme with massless $b$-quarks, which makes an IR-safe flavour-aware jet algorithm indispensable.

\subsection{$pp \to Z/\gamma^*(\to \ell\bar{\ell}) + b$-jet}\label{sec:pheno_ppzb}

We impose the following selection cuts:
\begin{itemize}
  \item $p_T(j_b) > 30$ GeV, $|\eta(j_b)| < 2.4$, $R = 0.5$,
  \item $p_T(\ell) > 20$ GeV, $|\eta(\ell)| < 2.4$,
  \item $71\,\text{GeV} \leq m(\ell\bar{\ell}) \leq 111\,\text{GeV}$,
  \item $\Delta R(j_b,\ell) > 0.5$,
\end{itemize}
which correspond to those assumed in the theoretical analysis of the process provided in Ref.~\cite{Gauld:2020deh}. The observables presented are the hardest $b$-jet's rapidity, $\eta(b_1)$, and transverse momentum, $p_T(b_1)$. Jets are clustered using the flavoured anti-$k_T$ jet algorithm, the flavoured $k_T$ jet algorithm up to NNLO and the standard anti-$k_T$ jet algorithm up to NLO. In all cases, we assign flavour according to the difference between the number of flavoured quarks and flavoured anti-quarks in the jet. We set the renormalisation and factorisation scale to the transverse mass of the lepton pair (equivalently, the transverse mass of the intermediate gauge boson), $\mu_R = \mu_F = m_T(\ell \bar{\ell}) = \sqrt{M_Z^2 + (\bm{p}_{T,\ell} + \bm{p}_{T,\bar{\ell}})^2}$.

The considered process is an excellent laboratory to investigate the impact of the flavoured anti-$k_T$ jet algorithm's parameter $a$ on physical observables. The $a$-parameter provides a measure of softness for flavour with respect to other scales in the process, notably $k_{T,\text{max}}$. Clearly, in the limit $a \to 0$ the IR-safety of the algorithm is lifted. This implies that in case of small but non-zero $a$-values, IR-sensitive observables are dominated by large logarithms of $a$, which spoils perturbative convergence. On the other hand, a small value of $a$ is in principle preferred because it leads to a smaller modification of the standard anti-$k_T$ jet algorithm. Conversely, a large $a$-value protects from large logarithms, but leads to large modifications of the clustering. The optimal value of $a$ lies in-between these two extremes and must be determined with a dedicated analysis. For comparison, we also consider the flavoured $k_T$ algorithm with $\alpha=2$ and the beam distance Eq.~\eqref{eq:flavourkt_beam} including the lepton momenta in the sum.

\begin{figure}[t]
 \centering
 \includegraphics[page=1,width=0.45\textwidth]{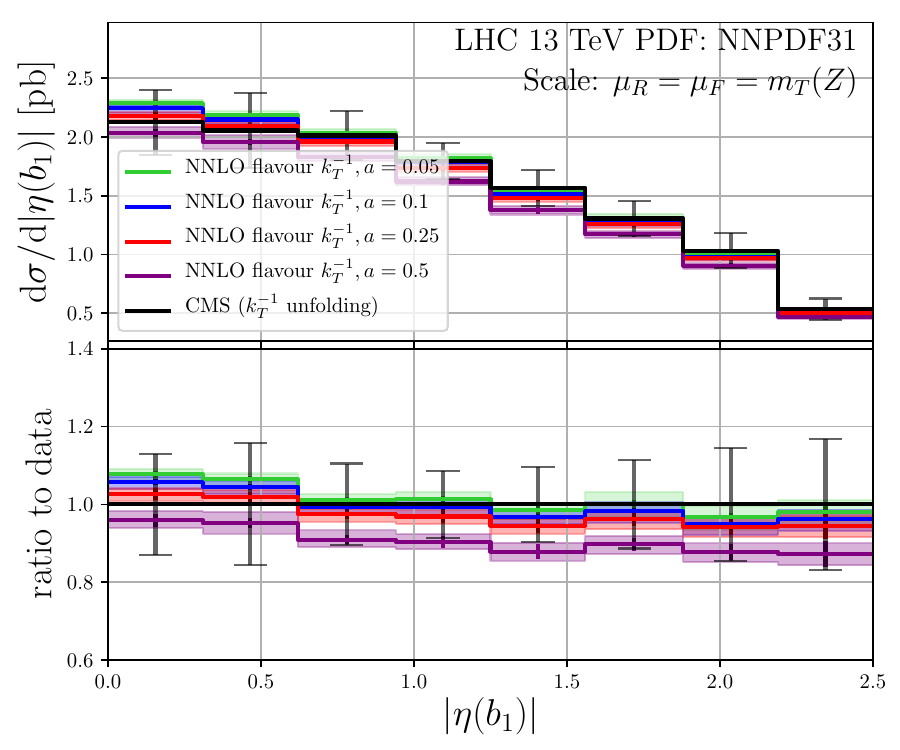}
 \includegraphics[page=2,width=0.45\textwidth]{./figs/ppzb/ppzb_NNLOvs_data.pdf}
 \caption{Comparison of differential distributions of the hardest jet's rapidity (left panel) and transverse momentum (right panel) for the process $pp \to Z/\gamma^*(\to \ell\bar{\ell}) + b$-jet obtained using different jet algorithms at NNLO accuracy in the phase space defined at the beginning of Section~\ref{sec:pheno_ppzb}. Also shown are measurement results by the CMS collaboration \cite{CMS:2016gmz}.}
 \label{fig:zb_1}
\end{figure}

\begin{figure}[t]
 \centering
 \includegraphics[page=1,width=0.45\textwidth]{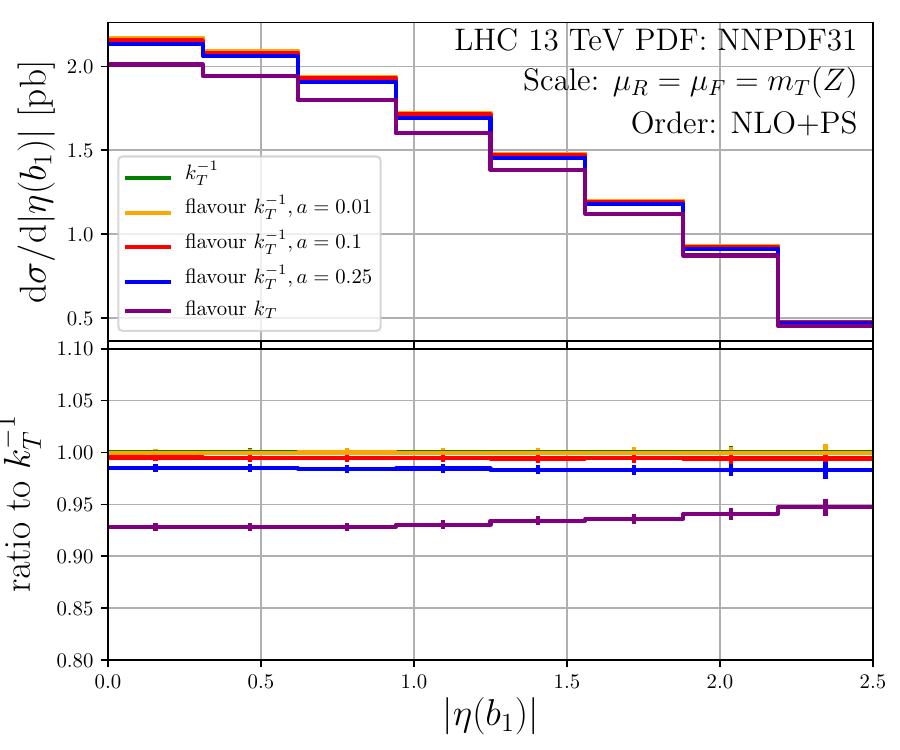}
 \includegraphics[page=2,width=0.45\textwidth]{./figs/ppzb/ppzb_algorithms_NLO+PS.pdf}
 \caption{Same as Fig.~\ref{fig:zb_1} but with NLO+PS (parton shower simulation matched at NLO QCD) accuracy in the phase space defined at the beginning of Section~\ref{sec:pheno_ppzb}.}
 \label{fig:zb_2}
\end{figure}

In Fig.~\ref{fig:zb_1}, we demonstrate the $a$-parameter dependence of the hardest $b$-jet rapidity and transverse momentum distributions at NNLO QCD for moderate to large values $a = 0.1, 0.25, 0.5$. The largest value yields distributions that are closest to those obtained with the flavoured $k_T$ algorithm. For the smallest value of $a$, the difference between the distributions amounts to about 5\% for most of the plotted range. We are, of course, more interested, in the difference between the flavoured anti-$k_T$ and the standard anti-$k_T$ algorithm. This cannot be studied at NNLO QCD, since the standard algorithm is IR safe only through NLO, but we can still employ a parton shower event-generator matched at NLO in QCD. We use \textsc{MadGraph5\textunderscore aMC@NLO} \cite{Alwall:2014hca} v.3.1.1 and refer to the results as NLO+PS. We use $n_f = 5$ massless quark flavours for the fixed-order NLO component, which is matched to the Pythia shower, which introduces massive quark flavours. The hadronisation stage is switched off. As expected, the NLO+PS distributions obtained with the new algorithm shown in Fig.~\ref{fig:zb_2} are closest to those of the standard anti-$k_T$ algorithm for the smallest value of $a$, while for the largest value of $a$ the differences between the distributions obtained with these two algorithms amount to about 5\%. This is consistent with the 10\% difference between distributions obtained with the flavoured $k_T$ and the standard anti-$k_T$ algorithm. Of course, this difference might be reduced or increased by varying the $\alpha$-parameter and beam distance definition of the flavoured $k_T$ algorithm. We did not perform any tuning of these parameters in order to minimize the difference between the results obtained with the two algorithms.

\begin{figure}[t]
 \centering
 \includegraphics[page=1,width=0.45\textwidth]{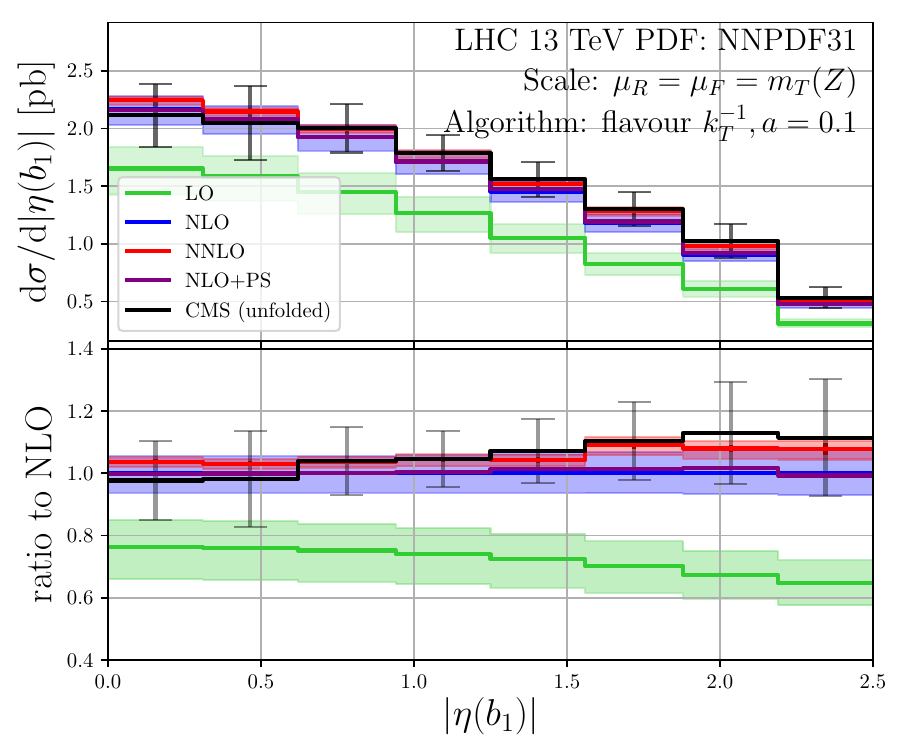}
 \includegraphics[page=2,width=0.45\textwidth]{./figs/ppzb/ppzb_flakt_0100.pdf}

 \includegraphics[page=1,width=0.45\textwidth]{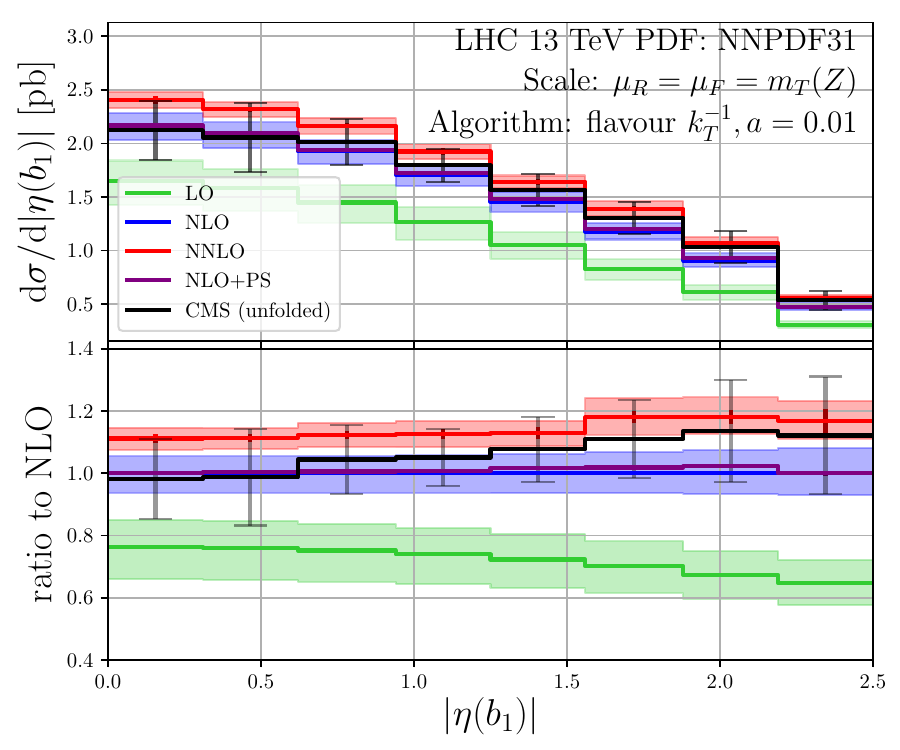}
 \includegraphics[page=2,width=0.45\textwidth]{./figs/ppzb/ppzb_flakt_0010.pdf}

 \includegraphics[page=1,width=0.45\textwidth]{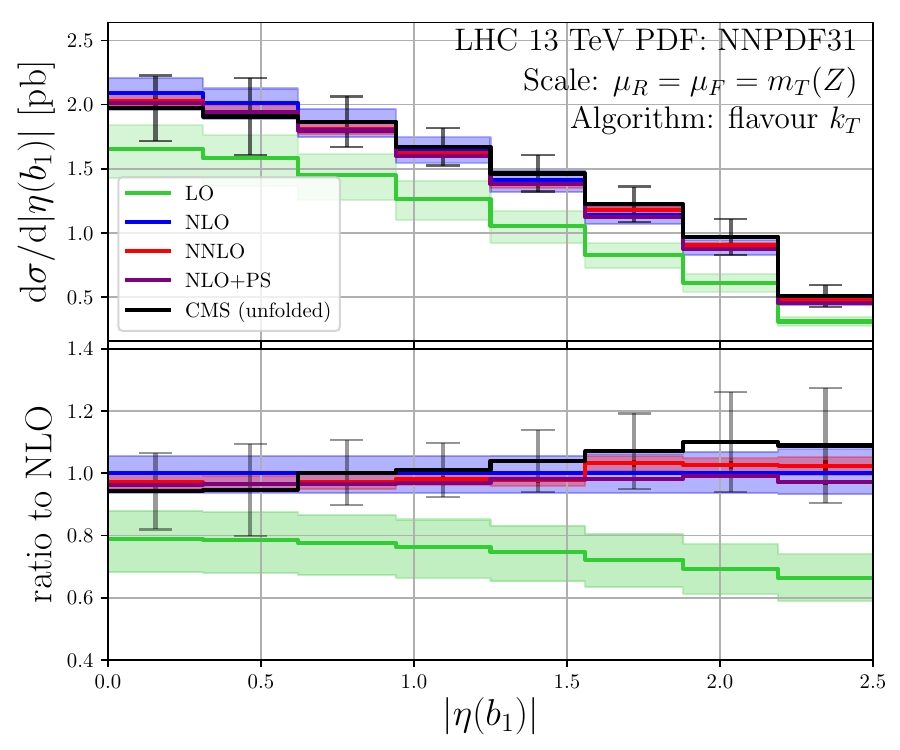}
 \includegraphics[page=2,width=0.45\textwidth]{./figs/ppzb/ppzb_flkt.pdf}
 \caption{Same as Fig.~\ref{fig:zb_1} but at LO, NLO, NNLO and NLO+PS accuracy in the phase space defined at the beginning of Section~\ref{sec:pheno_ppzb}. Also shown are measurement results by the CMS collaboration \cite{CMS:2016gmz}.}
 \label{fig:zb_3}
\end{figure}

In order to study the influence of the $a$-parameter on perturbative convergence, we plot in Fig.~\ref{fig:zb_3} the hardest $b$-jet's rapidity and transverse momentum distributions at LO, NLO, NNLO, NLO+PS for $a = 0.01$ and $a = 0.1$. We also provide plots for the flavoured $k_T$ algorithm, and superimpose the measurement results obtained by the CMS collaboration \cite{CMS:2016gmz} in each case. The data has been corrected by a bin-by-bin correction (indicated by the "unfolded" label) as described in Appendix~\ref{app:unfold}. Clearly, a value of $a = 0.01$ shows no satisfactory convergence in the case of the $b$-jet rapidity distribution, as the theory error estimate bands do not overlap at higher orders. For $a = 0.1$, we observe a convergent behaviour over the full rapidity range. The $b$-jet transverse momentum distribution is less sensitive to changes of $a$.

The conclusion of this study is that a reasonable value of the flavoured anti-$k_T$ jet algorithm parameter $a$ for the considered process and distributions is $a = 0.1$.

\subsection{Top-quark pair production and decay}\label{sec:pheno_pptt}

As a second example, we study $b$-jets in the context of the production and decay of a top-quark pair in the di-lepton channel within the Narrow Width Approximation \cite{Behring:2019iiv,Czakon:2020qbd}. We mimic the fiducial phase space cuts used in the experimental measurement of the CMS collaboration described in Ref.~\cite{CMS:2018adi}. In the context of our calculation at NNLO in QCD, events are selected by requiring \cite{Czakon:2020qbd}:
\begin{itemize}
\item $p_T(\ell) \geq 20 \;\text{GeV}$ and $|\eta(\ell)| \leq 2.4$ for both charged leptons\,,
\item $m(\ell\bar{\ell}) \geq 20\; \text{GeV}$\,,
\item two identified $b$-jets with $p_T \geq 30 \;\text{GeV}$, $|y| \leq 2.4$, $R = 0.4$, separated from the leptons by $\Delta R (j_b,\ell) \geq 0.4$.
\end{itemize}
The analysis presented in Ref.~\cite{Czakon:2020qbd} used the standard anti-$k_T$ algorithm. In that reference, it has been argued that the effect of soft $b$-quark-anti-quark pairs, if treated correctly, must be numerically small. The dangerous double-real contributions with final state $\ell \bar{\ell} \nu \bar{\nu} b\bar{b} + b\bar{b}$ have been included assuming a lower cutoff on the emission energy of $b\bar{b}$ pairs yielding numerically finite cross sections. While by no means ideal, this method allowed to achieve perturbatively stable and physically sensible results.

\begin{figure}[t]
 \centering
 \includegraphics[width=0.31\textwidth]{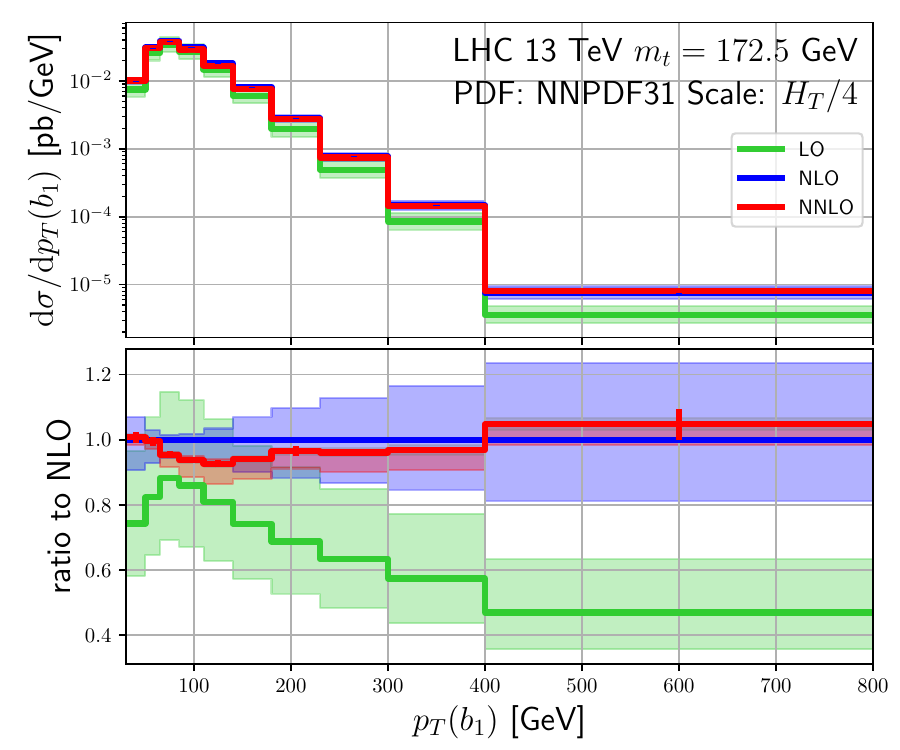}
 \includegraphics[width=0.31\textwidth]{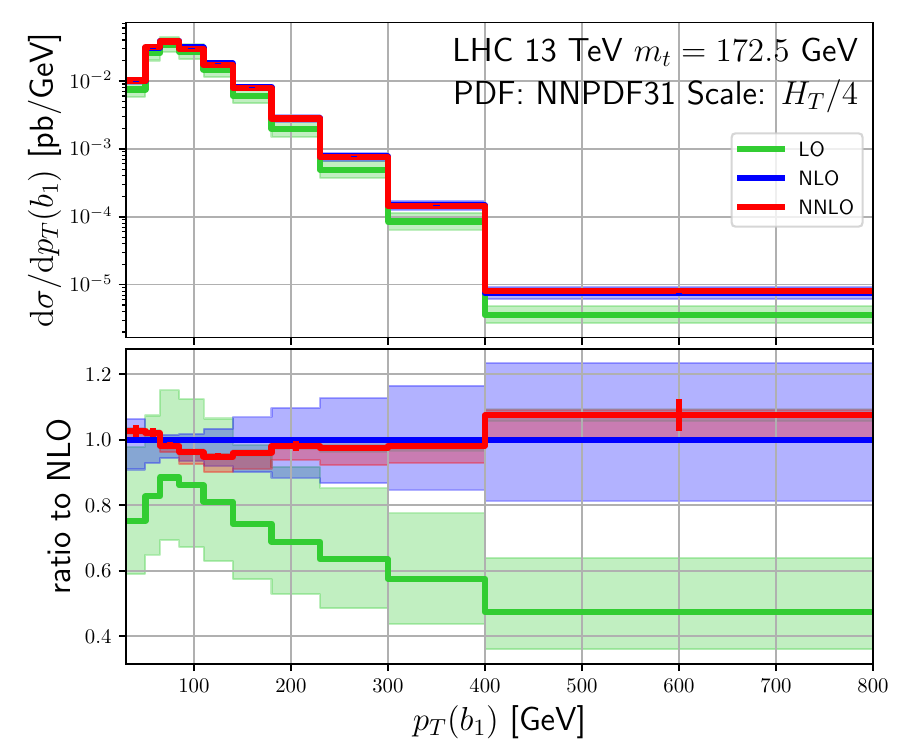}
 \includegraphics[width=0.31\textwidth]{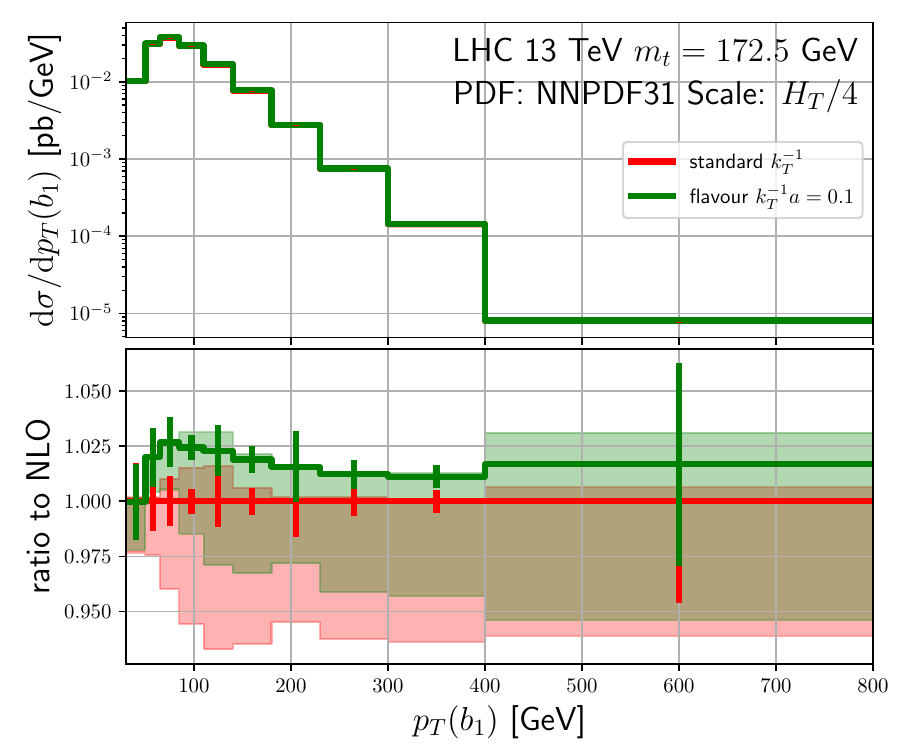}
 \caption{Differential distributions of the hardest $b$-jet transverse momentum in the process $pp \to t\bar{t} \to \ell\bar{\ell}\nu\bar{\nu} + 2 b$-jets, obtained at LO, NLO and NNLO in QCD with the flavoured anti-$k_T$ jet algorithm (left panel) and using the standard anti-$k_T$ jet algorithm with a lower cutoff on the energy of $b\bar{b}$ pairs \cite{Czakon:2020qbd} (middle panel). The results are compared at NNLO QCD in the right panel.}
 \label{fig:ttbar_1}
\end{figure}

\begin{figure}[t]
 \centering
 \includegraphics[width=0.31\textwidth]{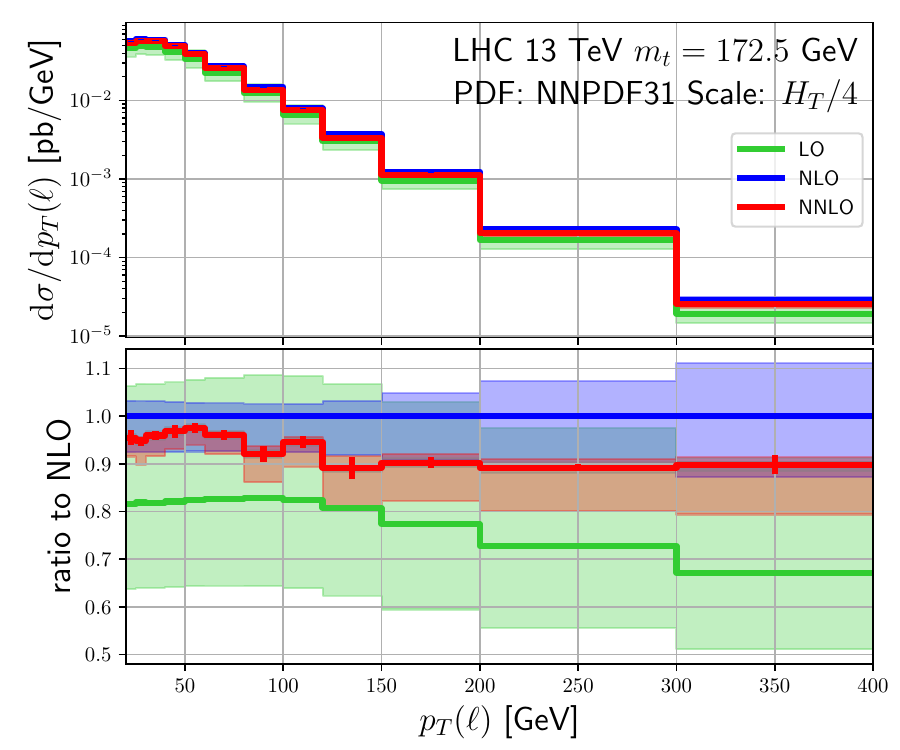}
 \includegraphics[width=0.31\textwidth]{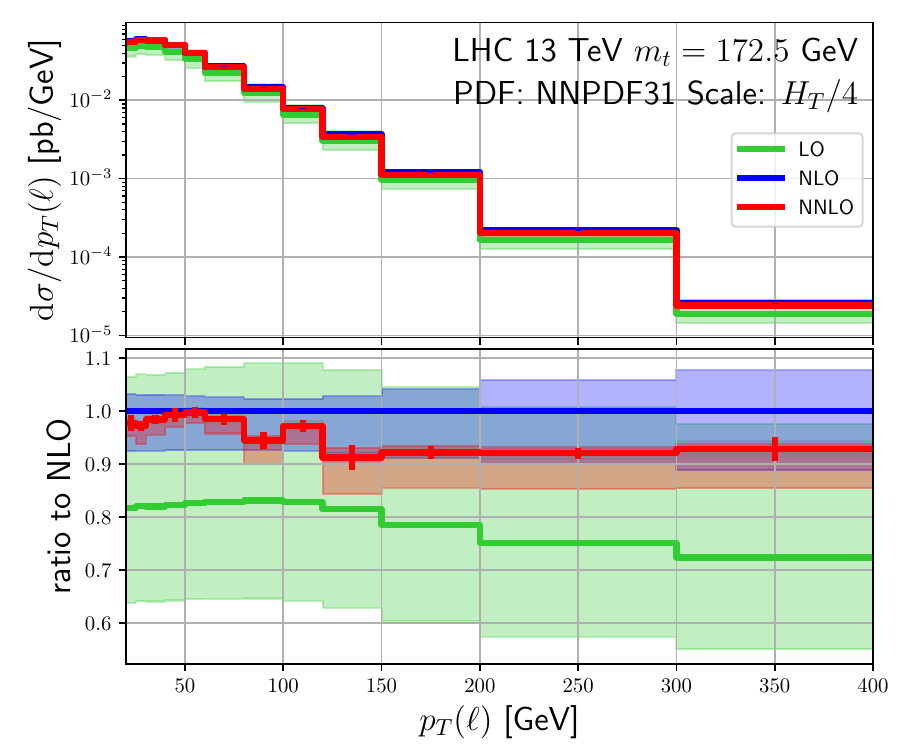}
 \includegraphics[width=0.31\textwidth]{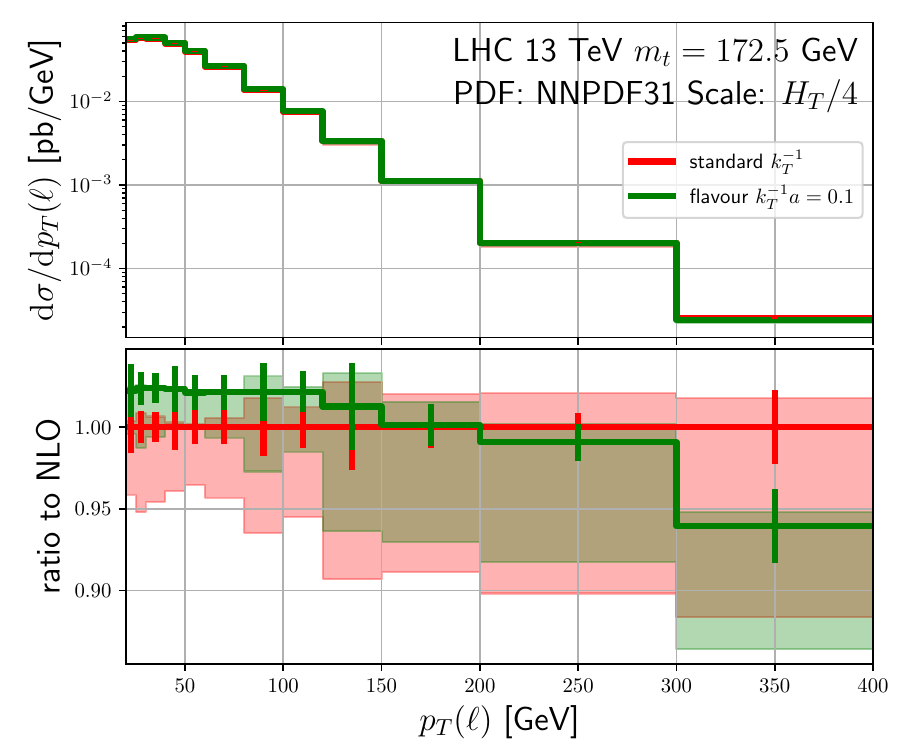}
 \caption{Same as Fig.~\ref{fig:ttbar_1} but for the transverse momentum of the lepton.}
 \label{fig:ttbar_2}
\end{figure}

\begin{figure}[t]
 \centering
 \includegraphics[width=0.31\textwidth]{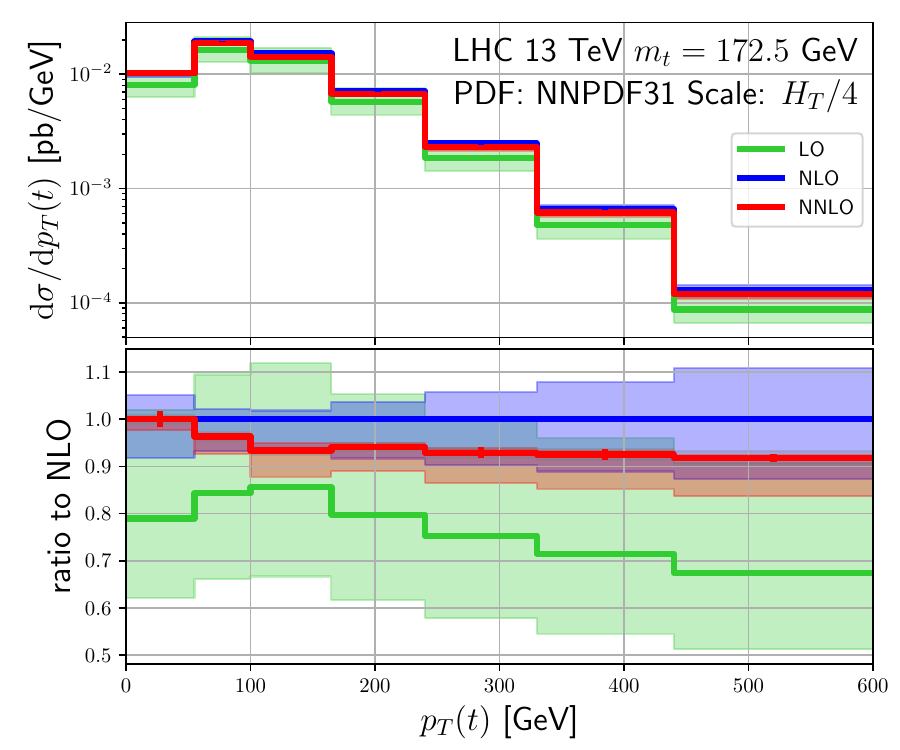}
 \includegraphics[width=0.31\textwidth]{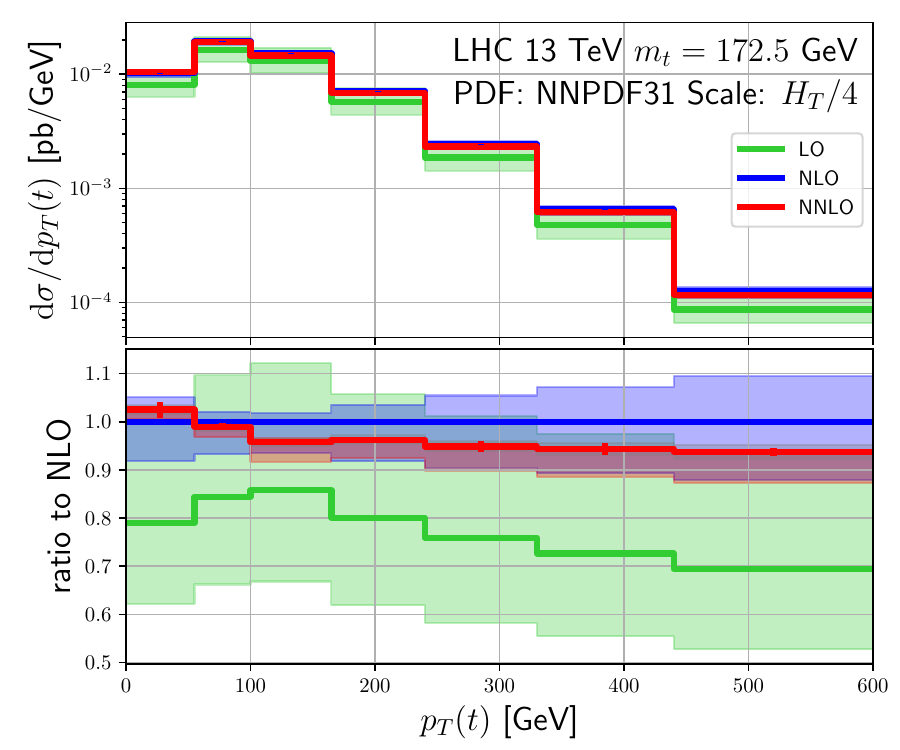}
 \includegraphics[width=0.31\textwidth]{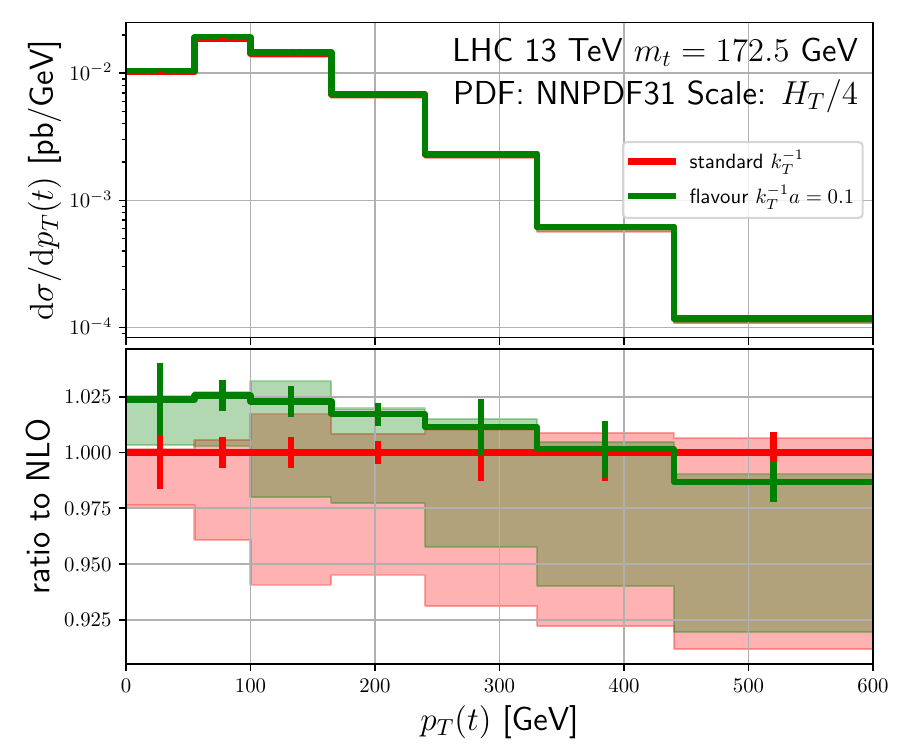}
 \caption{Same as Fig.~\ref{fig:ttbar_1} but for the transverse momentum of the reconstructed top quark.}
 \label{fig:ttbar_3}
\end{figure}

Here, we would like to compare the results of Ref.~\cite{Czakon:2020qbd} with the corresponding results obtained with our IR-safe flavoured anti-$k_T$ algorithm. In Fig.~\ref{fig:ttbar_1}, we show the transverse momentum of the leading $b$-jet computed with the cutoff method on the left and the IR-safe algorithm in the middle panel. The right panel shows a direct comparison of the results at NNLO. The differences between the results are roughly $\sim 1\%$, generally smaller than the scale uncertainty. The perturbative corrections are very similar in both cases as they are driven by partonic channels which are not sensitive to the modification of the algorithm. An analogous behaviour transpires for other observables which only indirectly depend on the $b$-jet definition, as for example the charged lepton transverse momentum, see Fig.~\ref{fig:ttbar_2}, or the reconstructed top-quark momentum, see Fig.~\ref{fig:ttbar_3}.

\section{Conclusions and outlook}\label{sec:conclusion}

In this work, we have presented the flavoured anti-$k_T$ jet algorithm - a minimal modification of the standard anti-$k_T$ algorithm that is flavour aware, infrared safe and can be easily implemented in perturbative fixed-order cross section calculations. Our construction is a viable alternative to the established flavoured $k_T$ algorithm, although it still requires the determination of unfolding corrections when performing precise comparisons to data.

The new algorithm contains a damping function, $\mathcal{S}_{ij}$, with specific properties. We have given a particular definition, Eq.~\eqref{eq:Sij}, of this function that depends on a ``softness parameter'' $a$. A study of the process $pp \to Z/\gamma^*(\to \ell\bar{\ell}) + b$-jet leads to a recommended value for this parameter, $a = 0.1$. However, the preferred $a$-value may be application dependent.

Besides the parameter $a$, the definition of the function $\mathcal{S}_{ij}$, Eq.~\eqref{eq:Sij}, depends on a dimensionful scale, $k_{T,\text{max}}$, which we have chosen to be equal the to hardest pseudo-jet's transverse momentum in a recombination step. Other choices are possible here as well. For example, in the case of the top-quark pair production process, it could be taken equal to the top-quark mass, or even the renormalization scale as long as the latter is not jet based.

The tuning of the above details of the algorithm is left to future studies.

\subsection*{Note added:} After the preprint version of this article has been published on arXiv, there appeared another proposal for a flavour-aware jet algorithm \cite{Gauld:2022lem}.

\begin{acknowledgments}
R.P. would like to thank Gavin Salam and Fabrizio Caola for a discussion on the subtleties of IR safety.

The work of M.C. was supported by the Deutsche Forschungsgemeinschaft under grant 396021762 – TRR 257. The research of A.M. and R.P. has received funding from the European Research Council (ERC) under the European Union’s Horizon 2020 Research and Innovation Programme (grant agreement no. 683211). A.M. was also supported by the UK STFC grants ST/L002760/1 and ST/K004883/1. R.P. acknowledges support from the Leverhulme Trust and the Isaac Newton Trust.
\end{acknowledgments}

\appendix

\section{Unfolding of Z+b-jet data}\label{app:unfold}

\begin{figure}[t]
 \centering
 \includegraphics[page=1,width=0.45\textwidth]{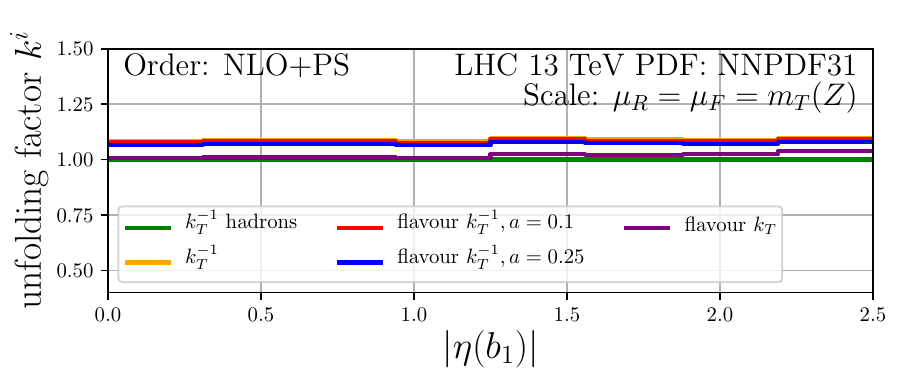}%
 \includegraphics[page=2,width=0.45\textwidth]{./figs/ppzb/ppzb_algorithms_NLO+PS_unfold.pdf}
 \caption{Unfolding factors $k^i$ for $|\eta(b_1)|$ and $p_T(b_1)$.}
 \label{fig:app_unfold}
\end{figure}

Although we have indicated measurement results in Figs.~\ref{fig:zb_1} and \ref{fig:zb_3}, a meaningful comparison between theory and data actually requires an unfolding as explained in the Introduction. In this Appendix, we present unfolding corrections for cross sections differential in $|\eta(b_1)|$ and $p_T(b_1)$. To this end, we evaluate the 
distributions using an NLO+PS simulation including hadronisation and experimental flavour tagging, as well as using a NLO+PS simulation without the hadronisation phase and with clustering and flavour tagging according to the algorithms considered in the text. We use the same \textsc{MadGraph5\textunderscore aMC@NLO} setup as in Section~\ref{sec:pheno_ppzb}. Events with hadronisation enabled are analysed with the \textsc{Rivet} routine provided together with the data from Ref.~\cite{CMS:2016gmz}. In Figure \ref{fig:app_unfold}, we show the differential $k$-factors defined as follows:
\begin{equation}
    k^i(x) = \frac{\dd \sigma^i(x)}{\dd \sigma^{\text{had}}(x)} \; .
\end{equation}
Here, the numerator $\dd \sigma^i(x)$ corresponds to the algorithm $i$. This correction does not contain bin-to-bin correlations which however are expected to be small, see Ref.~\cite{Gauld:2020deh}. In the case of the leading $b$-jet rapidity distribution, anti-$k_T$ algorithms, both flavoured and non-flavoured, require a flat correction of about $8\%$. The flavoured $k_T$ algorithm does not require any correction within the statistics of the calculation. For the leading $b$-jet transverse momentum, the corrections required by the anti-$k_T$ algorithms coincide again, but are not flat anymore and vary from about $+8\%$ at low $p_T$ to $-8\%$ at $p_T$ of the order of 300 GeV. The unfolding correction of the flavoured $k_T$ algorithm is negligible for low transverse momenta, but increases to twice the size of that of anti-$k_T$ algorithms for large transverse momenta.
 
\bibliography{main}

\providecommand{\href}[2]{#2}\begingroup\raggedright\begin{thebibliography}{10}

\bibitem{Salam:2010nqg}
G.~P. Salam, \emph{{Towards Jetography}},
  \href{https://doi.org/10.1140/epjc/s10052-010-1314-6}{\emph{Eur. Phys. J. C}
  {\bfseries 67} (2010) 637} [\href{https://arxiv.org/abs/0906.1833}{{\ttfamily
  0906.1833}}].

\bibitem{Larkoski:2017jix}
A.~J. Larkoski, I.~Moult and B.~Nachman, \emph{{Jet Substructure at the Large
  Hadron Collider: A Review of Recent Advances in Theory and Machine
  Learning}}, \href{https://doi.org/10.1016/j.physrep.2019.11.001}{\emph{Phys.
  Rept.} {\bfseries 841} (2020) 1}
  [\href{https://arxiv.org/abs/1709.04464}{{\ttfamily 1709.04464}}].

\bibitem{Marzani:2019hun}
S.~Marzani, G.~Soyez and M.~Spannowsky, \emph{{Looking inside jets: an
  introduction to jet substructure and boosted-object phenomenology}},
  vol.~958. Springer, 2019,
  \href{https://doi.org/10.1007/978-3-030-15709-8}{10.1007/978-3-030-15709-8},
  [\href{https://arxiv.org/abs/1901.10342}{{\ttfamily 1901.10342}}].

\bibitem{Gras:2017jty}
P.~Gras, S.~H\"oche, D.~Kar, A.~Larkoski, L.~L\"onnblad, S.~Pl\"atzer et~al.,
  \emph{{Systematics of quark/gluon tagging}},
  \href{https://doi.org/10.1007/JHEP07(2017)091}{\emph{JHEP} {\bfseries 07}
  (2017) 091} [\href{https://arxiv.org/abs/1704.03878}{{\ttfamily
  1704.03878}}].

\bibitem{Banfi:2006hf}
A.~Banfi, G.~P. Salam and G.~Zanderighi, \emph{{Infrared safe definition of jet
  flavor}}, \href{https://doi.org/10.1140/epjc/s2006-02552-4}{\emph{Eur. Phys.
  J. C} {\bfseries 47} (2006) 113}
  [\href{https://arxiv.org/abs/hep-ph/0601139}{{\ttfamily hep-ph/0601139}}].

\bibitem{Banfi:2007gu}
A.~Banfi, G.~P. Salam and G.~Zanderighi, \emph{{Accurate QCD predictions for
  heavy-quark jets at the Tevatron and LHC}},
  \href{https://doi.org/10.1088/1126-6708/2007/07/026}{\emph{JHEP} {\bfseries
  07} (2007) 026} [\href{https://arxiv.org/abs/0704.2999}{{\ttfamily
  0704.2999}}].

\bibitem{Banfi:2010xy}
A.~Banfi, G.~P. Salam and G.~Zanderighi, \emph{{Phenomenology of event shapes
  at hadron colliders}},
  \href{https://doi.org/10.1007/JHEP06(2010)038}{\emph{JHEP} {\bfseries 06}
  (2010) 038} [\href{https://arxiv.org/abs/1001.4082}{{\ttfamily 1001.4082}}].

\bibitem{Banfi:2016zlc}
A.~Banfi, H.~McAslan, P.~F. Monni and G.~Zanderighi, \emph{{The two-jet rate in
  $e^+e^-$ at next-to-next-to-leading-logarithmic order}},
  \href{https://doi.org/10.1103/PhysRevLett.117.172001}{\emph{Phys. Rev. Lett.}
  {\bfseries 117} (2016) 172001}
  [\href{https://arxiv.org/abs/1607.03111}{{\ttfamily 1607.03111}}].

\bibitem{Weinzierl:2006yt}
S.~Weinzierl, \emph{{The Forward-backward asymmetry at NNLO revisited}},
  \href{https://doi.org/10.1016/j.physletb.2006.11.076}{\emph{Phys. Lett. B}
  {\bfseries 644} (2007) 331}
  [\href{https://arxiv.org/abs/hep-ph/0609021}{{\ttfamily hep-ph/0609021}}].

\bibitem{Caola:2014daa}
F.~Caola, A.~Czarnecki, Y.~Liang, K.~Melnikov and R.~Szafron, \emph{{Muon decay
  spin asymmetry}},
  \href{https://doi.org/10.1103/PhysRevD.90.053004}{\emph{Phys. Rev. D}
  {\bfseries 90} (2014) 053004}
  [\href{https://arxiv.org/abs/1403.3386}{{\ttfamily 1403.3386}}].

\bibitem{Caola:2017xuq}
F.~Caola, G.~Luisoni, K.~Melnikov and R.~R\"ontsch, \emph{{NNLO QCD corrections
  to associated $WH$ production and $H \to b \bar b$ decay}},
  \href{https://doi.org/10.1103/PhysRevD.97.074022}{\emph{Phys. Rev. D}
  {\bfseries 97} (2018) 074022}
  [\href{https://arxiv.org/abs/1712.06954}{{\ttfamily 1712.06954}}].

\bibitem{Ferrera:2017zex}
G.~Ferrera, G.~Somogyi and F.~Tramontano, \emph{{Associated production of a
  Higgs boson decaying into bottom quarks at the LHC in full NNLO QCD}},
  \href{https://doi.org/10.1016/j.physletb.2018.03.021}{\emph{Phys. Lett. B}
  {\bfseries 780} (2018) 346}
  [\href{https://arxiv.org/abs/1705.10304}{{\ttfamily 1705.10304}}].

\bibitem{Gauld:2019yng}
R.~Gauld, A.~Gehrmann-De~Ridder, E.~W.~N. Glover, A.~Huss and I.~Majer,
  \emph{{Associated production of a Higgs boson decaying into bottom quarks and
  a weak vector boson decaying leptonically at NNLO in QCD}},
  \href{https://doi.org/10.1007/JHEP10(2019)002}{\emph{JHEP} {\bfseries 10}
  (2019) 002} [\href{https://arxiv.org/abs/1907.05836}{{\ttfamily
  1907.05836}}].

\bibitem{Gauld:2021ule}
R.~Gauld, A.~Gehrmann-De~Ridder, E.~W.~N. Glover, A.~Huss and I.~Majer,
  \emph{{VH + jet production in hadron-hadron collisions up to order $
  {\alpha}_{\mathrm{s}}^3 $ in perturbative QCD}},
  \href{https://doi.org/10.1007/JHEP03(2022)008}{\emph{JHEP} {\bfseries 03}
  (2022) 008} [\href{https://arxiv.org/abs/2110.12992}{{\ttfamily
  2110.12992}}].

\bibitem{Behring:2020uzq}
A.~Behring, W.~Bizo\'n, F.~Caola, K.~Melnikov and R.~R\"ontsch, \emph{{Bottom
  quark mass effects in associated $WH$ production with the $H \to b\bar{b}$
  decay through NNLO QCD}},
  \href{https://doi.org/10.1103/PhysRevD.101.114012}{\emph{Phys. Rev. D}
  {\bfseries 101} (2020) 114012}
  [\href{https://arxiv.org/abs/2003.08321}{{\ttfamily 2003.08321}}].

\bibitem{Gauld:2020deh}
R.~Gauld, A.~Gehrmann-De~Ridder, E.~W.~N. Glover, A.~Huss and I.~Majer,
  \emph{{Predictions for $Z$ -Boson Production in Association with a $b$-Jet at
  $\mathcal {O}(\alpha_s^3)$}},
  \href{https://doi.org/10.1103/PhysRevLett.125.222002}{\emph{Phys. Rev. Lett.}
  {\bfseries 125} (2020) 222002}
  [\href{https://arxiv.org/abs/2005.03016}{{\ttfamily 2005.03016}}].

\bibitem{Czakon:2020coa}
M.~Czakon, A.~Mitov, M.~Pellen and R.~Poncelet, \emph{{NNLO QCD predictions for
  W+c-jet production at the LHC}},
  \href{https://doi.org/10.1007/JHEP06(2021)100}{\emph{JHEP} {\bfseries 06}
  (2021) 100} [\href{https://arxiv.org/abs/2011.01011}{{\ttfamily
  2011.01011}}].

\bibitem{Hartanto:2022qhh}
H.~B. Hartanto, R.~Poncelet, A.~Popescu and S.~Zoia, \emph{{NNLO QCD
  corrections to $Wb\bar{b}$ production at the LHC}},
  \href{https://arxiv.org/abs/2205.01687}{{\ttfamily 2205.01687}}.

\bibitem{Czakon:2021ohs}
M.~L. Czakon, T.~Generet, A.~Mitov and R.~Poncelet, \emph{{B-hadron production
  in NNLO QCD: application to LHC t$ \overline{t} $ events with leptonic
  decays}}, \href{https://doi.org/10.1007/JHEP10(2021)216}{\emph{JHEP}
  {\bfseries 10} (2021) 216}
  [\href{https://arxiv.org/abs/2102.08267}{{\ttfamily 2102.08267}}].

\bibitem{Anger:2017glm}
F.~R. Anger, F.~Febres~Cordero, H.~Ita and V.~Sotnikov, \emph{{NLO QCD
  predictions for $Wb\bar b$ production in association with up to three light
  jets at the LHC}},
  \href{https://doi.org/10.1103/PhysRevD.97.036018}{\emph{Phys. Rev. D}
  {\bfseries 97} (2018) 036018}
  [\href{https://arxiv.org/abs/1712.05721}{{\ttfamily 1712.05721}}].

\bibitem{Astill:2018ivh}
W.~Astill, W.~Bizo\'n, E.~Re and G.~Zanderighi, \emph{{NNLOPS accurate
  associated HZ production with $ H\to b\overline{b} $ decay at NLO}},
  \href{https://doi.org/10.1007/JHEP11(2018)157}{\emph{JHEP} {\bfseries 11}
  (2018) 157} [\href{https://arxiv.org/abs/1804.08141}{{\ttfamily
  1804.08141}}].

\bibitem{Bizon:2019tfo}
W.~Bizo\'n, E.~Re and G.~Zanderighi, \emph{{NNLOPS description of the $H \to
  b\overline{b} $ decay with MiNLO}},
  \href{https://doi.org/10.1007/JHEP06(2020)006}{\emph{JHEP} {\bfseries 06}
  (2020) 006} [\href{https://arxiv.org/abs/1912.09982}{{\ttfamily
  1912.09982}}].

\bibitem{Zanoli:2021iyp}
S.~Zanoli, M.~Chiesa, E.~Re, M.~Wiesemann and G.~Zanderighi,
  \emph{{Next-to-next-to-leading order event generation for $VH$ production
  with $H\to b\bar{b}$ decay}},
  \href{https://arxiv.org/abs/2112.04168}{{\ttfamily 2112.04168}}.

\bibitem{Bevilacqua:2021tzp}
G.~Bevilacqua, H.~Y. Bi, F.~Febres~Cordero, H.~B. Hartanto, M.~Kraus, J.~Nasufi
  et~al., \emph{{Modeling uncertainties of $t\bar{t}W^\pm$ multilepton
  signatures}}, \href{https://doi.org/10.1103/PhysRevD.105.014018}{\emph{Phys.
  Rev. D} {\bfseries 105} (2022) 014018}
  [\href{https://arxiv.org/abs/2109.15181}{{\ttfamily 2109.15181}}].

\bibitem{Larkoski:2013eya}
A.~J. Larkoski, G.~P. Salam and J.~Thaler, \emph{{Energy Correlation Functions
  for Jet Substructure}},
  \href{https://doi.org/10.1007/JHEP06(2013)108}{\emph{JHEP} {\bfseries 06}
  (2013) 108} [\href{https://arxiv.org/abs/1305.0007}{{\ttfamily 1305.0007}}].

\bibitem{Baberuxki:2019ifp}
N.~Baberuxki, C.~T. Preuss, D.~Reichelt and S.~Schumann, \emph{{Resummed
  predictions for jet-resolution scales in multijet production in $e^+e^-$
  annihilation}}, \href{https://doi.org/10.1007/JHEP04(2020)112}{\emph{JHEP}
  {\bfseries 04} (2020) 112}
  [\href{https://arxiv.org/abs/1912.09396}{{\ttfamily 1912.09396}}].

\bibitem{Cacciari:2008gp}
M.~Cacciari, G.~P. Salam and G.~Soyez, \emph{{The anti-$k_t$ jet clustering
  algorithm}}, \href{https://doi.org/10.1088/1126-6708/2008/04/063}{\emph{JHEP}
  {\bfseries 04} (2008) 063} [\href{https://arxiv.org/abs/0802.1189}{{\ttfamily
  0802.1189}}].

\bibitem{Heinrich:2013qaa}
G.~Heinrich, A.~Maier, R.~Nisius, J.~Schlenk and J.~Winter, \emph{{NLO QCD
  corrections to $W^{+} W^{-}b\bar{b}$ production with leptonic decays in the
  light of top quark mass and asymmetry measurements}},
  \href{https://doi.org/10.1007/JHEP06(2014)158}{\emph{JHEP} {\bfseries 06}
  (2014) 158} [\href{https://arxiv.org/abs/1312.6659}{{\ttfamily 1312.6659}}].

\bibitem{Caletti:2021oor}
S.~Caletti, O.~Fedkevych, S.~Marzani, D.~Reichelt, S.~Schumann, G.~Soyez
  et~al., \emph{{Jet angularities in Z+jet production at the LHC}},
  \href{https://doi.org/10.1007/JHEP07(2021)076}{\emph{JHEP} {\bfseries 07}
  (2021) 076} [\href{https://arxiv.org/abs/2104.06920}{{\ttfamily
  2104.06920}}].

\bibitem{Buckley:2015gua}
A.~Buckley and C.~Pollard, \emph{{QCD-aware partonic jet clustering for
  truth-jet flavour labelling}},
  \href{https://doi.org/10.1140/epjc/s10052-016-3925-z}{\emph{Eur. Phys. J. C}
  {\bfseries 76} (2016) 71} [\href{https://arxiv.org/abs/1507.00508}{{\ttfamily
  1507.00508}}].

\bibitem{Fedkevych:2022mid}
O.~Fedkevych, C.~K. Khosa, S.~Marzani and F.~Sforza, \emph{{Identification of
  b-jets using QCD-inspired observables}},
  \href{https://arxiv.org/abs/2202.05082}{{\ttfamily 2202.05082}}.

\bibitem{Caletti:2022glq}
S.~Caletti, A.~J. Larkoski, S.~Marzani and D.~Reichelt, \emph{{A Fragmentation
  Approach to Jet Flavor}},  \href{https://arxiv.org/abs/2205.01117}{{\ttfamily
  2205.01117}}.

\bibitem{Caletti:2022hnc}
S.~Caletti, A.~J. Larkoski, S.~Marzani and D.~Reichelt, \emph{{Practical Jet
  Flavour Through NNLO}},  \href{https://arxiv.org/abs/2205.01109}{{\ttfamily
  2205.01109}}.

\bibitem{Gleisberg:2008ta}
T.~Gleisberg, S.~Hoeche, F.~Krauss, M.~Schonherr, S.~Schumann, F.~Siegert
  et~al., \emph{{Event generation with SHERPA 1.1}},
  \href{https://doi.org/10.1088/1126-6708/2009/02/007}{\emph{JHEP} {\bfseries
  02} (2009) 007} [\href{https://arxiv.org/abs/0811.4622}{{\ttfamily
  0811.4622}}].

\bibitem{CMS:2016gmz}
{\scshape CMS} collaboration, \emph{{Measurements of the associated production
  of a Z boson and b jets in pp collisions at ${\sqrt{s}} = 8\,\text {TeV} $}},
  \href{https://doi.org/10.1140/epjc/s10052-017-5140-y}{\emph{Eur. Phys. J. C}
  {\bfseries 77} (2017) 751}
  [\href{https://arxiv.org/abs/1611.06507}{{\ttfamily 1611.06507}}].

\bibitem{Alwall:2014hca}
J.~Alwall, R.~Frederix, S.~Frixione, V.~Hirschi, F.~Maltoni, O.~Mattelaer
  et~al., \emph{{The automated computation of tree-level and next-to-leading
  order differential cross sections, and their matching to parton shower
  simulations}}, \href{https://doi.org/10.1007/JHEP07(2014)079}{\emph{JHEP}
  {\bfseries 07} (2014) 079} [\href{https://arxiv.org/abs/1405.0301}{{\ttfamily
  1405.0301}}].

\bibitem{Behring:2019iiv}
A.~Behring, M.~Czakon, A.~Mitov, A.~S. Papanastasiou and R.~Poncelet,
  \emph{{Higher order corrections to spin correlations in top quark pair
  production at the LHC}},
  \href{https://doi.org/10.1103/PhysRevLett.123.082001}{\emph{Phys. Rev. Lett.}
  {\bfseries 123} (2019) 082001}
  [\href{https://arxiv.org/abs/1901.05407}{{\ttfamily 1901.05407}}].

\bibitem{Czakon:2020qbd}
M.~Czakon, A.~Mitov and R.~Poncelet, \emph{{NNLO QCD corrections to leptonic
  observables in top-quark pair production and decay}},
  \href{https://doi.org/10.1007/JHEP05(2021)212}{\emph{JHEP} {\bfseries 05}
  (2021) 212} [\href{https://arxiv.org/abs/2008.11133}{{\ttfamily
  2008.11133}}].

\bibitem{CMS:2018adi}
{\scshape CMS} collaboration, \emph{{Measurements of $\mathrm{t\overline{t}}$
  differential cross sections in proton-proton collisions at $\sqrt{s}=$ 13 TeV
  using events containing two leptons}},
  \href{https://doi.org/10.1007/JHEP02(2019)149}{\emph{JHEP} {\bfseries 02}
  (2019) 149} [\href{https://arxiv.org/abs/1811.06625}{{\ttfamily
  1811.06625}}].

\bibitem{Gauld:2022lem}
R.~Gauld, A.~Huss and G.~Stagnitto, \emph{{A dress of flavour to suit any
  jet}},  \href{https://arxiv.org/abs/2208.11138}{{\ttfamily 2208.11138}}.

\end{thebibliography}\endgroup
\bibliographystyle{JHEP}

\end{document}